\documentclass[letterpaper, 10 pt, conference]{ieeeconf}

\usepackage[utf8x]{inputenc}
\usepackage{cite}
\usepackage{amssymb,amsfonts,bm}
\usepackage{algorithmic}
\usepackage{graphicx}
\usepackage{textcomp}
\usepackage{amsmath,tikz}
\usepackage{color}
\usepackage{url}
\usepackage{mathtools}
\usepackage[hidelinks]{hyperref} 
\usepackage{rotating}
\usepackage{blkarray}
\usepackage{physics}
\usetikzlibrary{arrows}
\usepackage{booktabs}
\usepackage{multirow}
\let\proof\relax
\let\endproof\relax
\usepackage{amsthm}

\usepackage{caption}
\usepackage{subcaption}

\usepackage{etoolbox}
\let\classAND\AND
\let\AND\relax
\usepackage{algorithm}
\usepackage{algorithmic}

\let\AND\classAND
\AtBeginEnvironment{algorithmic}{\let\AND\algoAND}

\floatname{algorithm}{Algorithm}

\graphicspath{{eps/}{png/}}
\DeclareSymbolFont{symbolsC}{U}{pxsyc}{m}{n}
\DeclareMathSymbol{\coloneqq}{\mathrel}{symbolsC}{"42}

\newcommand{\vertiii}[1]{{\left\vert\kern-0.25ex\left\vert\kern-0.25ex\left\vert
		#1 
		\right\vert\kern-0.25ex\right\vert\kern-0.25ex\right\vert}}
\usepackage{multirow}
\usepackage{makecell}
\usepackage{adjustbox}
\usepackage{float}

\usepackage{pifont}
\usepackage{soul}

\IEEEoverridecommandlockouts

\begin{document}
	
\title{\LARGE\bfseries NeuralMUSIC: A Hybrid Neural–Subspace Framework for Robot Sound Source Localization}
\author{Yizhuo Yang, Junqiao Fan, Shenghai Yuan$^*$,~\IEEEmembership{Member,~IEEE}, Lihua Xie,~\IEEEmembership{Fellow,~IEEE}%
\thanks{Y. Yang, J. Fan and S. Yuan are with the School of Electrical and Electronic Engineering, Nanyang Technological University. Address: 50 Nanyang Avenue, Singapore 639798. 
L. Xie is with NTU–VinUni Joint Research Laboratory for Embodied AI and Robotics, School of Electrical and Electronic Engineering, Nanyang Technological University, Singapore 639798 and VinUniversity, Hanoi, Vietnam.
$^*$Corresponding author. Email: YIZHUO001@e.ntu.edu.sg \ \{shyuan,elxie\}@ntu.edu.sg}
\thanks{This work is supported by the National Research Foundation, Singapore, under its Medium-Sized Centre funding scheme through the Centre for Advanced Robotics Technology Innovation (CARTIN).} 
}

\maketitle

\begin{abstract}
Reliable sound source localization is fundamental to robot audition, enabling autonomous robots to perceive spatial cues and operate effectively in dynamic environments. 
Classical methods such as Multiple Signal Classification (MUSIC) offer strong theoretical foundations but degrade under low signal-to-noise ratios. 
While deep learning–based approaches achieve promising performance, they often struggle with limited generalization across conditions.
To address these challenges, we propose NeuralMUSIC, a hybrid neural–subspace framework for robotic sound source localization. Specifically, a neural network first estimates the spatial covariance matrix from multichannel microphone observations. The predicted covariance is then integrated into a classical MUSIC pipeline with eigenvalue decomposition (EVD) and pseudo-spectrum computation, followed by a Frequency Attention Fusion (FAF) module to produce the final DOA estimates.
To improve data efficiency, we further introduce a Self-supervised Spatial Correlation Learning (SSCL) strategy that leverages unlabeled acoustic data to capture spatial structure.
Extensive experiments across different robotic tasks demonstrate that NeuralMUSIC achieves competitive localization accuracy while exhibiting improved robustness and cross-domain generalization. 
The code is available at: \href{https://github.com/yizhuoyang/NeuralMUSIC.git}{https://github.com/yizhuoyang/NeuralMUSIC.git}.
\end{abstract}

\IEEEpeerreviewmaketitle

\section{Introduction}
\label{sec:intro}

Sound source localization (SSL) is a fundamental component of robot audition, providing robots with spatial awareness of their acoustic environment and enabling intelligent interaction with surrounding agents.
By estimating the Direction of Arrival (DOA) of sound sources, robots are able to effectively track active speakers \cite{masuyama2020self}, support human–robot interaction \cite{liu2025sound} and perceive dynamic agents in the surrounding environment \cite{yang2023av,yang2024kidnappable,schulz2021hearing}. 
However, achieving reliable DOA estimation in real-world robotic environments remains challenging due to noise, reverberation, multiple concurrent sound sources, and variations in acoustic environments and microphone configurations, which demand SSL methods with strong robustness and generalization capability.

Traditional SSL methods estimate the DOA by exploiting spatial correlations across microphone arrays. Representative approaches include cross-correlation-based delay estimation (e.g., GCC-PHAT \cite{liu2010continuous}), steered response power methods such as SRP-PHAT \cite{valin2004localization}, and subspace techniques like Multiple Signal Classification (MUSIC) \cite{schmidt1986multiple}, which construct the spatial covariance matrix and exploit its eigenstructure to separate signal and noise subspaces for high-resolution localization. Although effective in controlled settings, these classical methods rely on accurate array calibration and idealized signal assumptions. Their performance often degrades in low-SNR and reverberant environments commonly encountered by mobile robots operating in real-world settings.

With the rapid advancement of deep learning, data-driven approaches have been widely explored for SSL.
For example, Qayyum et al. \cite{qayyum2020DOANet} utilized an end-to-end 1D dilated CNN to regress azimuth and elevation from raw multichannel audio captured by a drone for search-and-rescue. Wu et al. \cite{wu2024afpild} employed CNN and GRU layers to extract spatial and temporal–spectral features from footstep sounds for pedestrian localization. 
These deep learning–based approaches improve robustness to noise and reverberation in complex acoustic environments. However, they often operate as black boxes with limited physical interpretability and exhibit poor generalization to unseen environments or different array geometries. These limitations hinder reliable deployment on robots operating in dynamic environments.

\begin{figure}[t]
    \centering
    \includegraphics[width=1.0\linewidth]{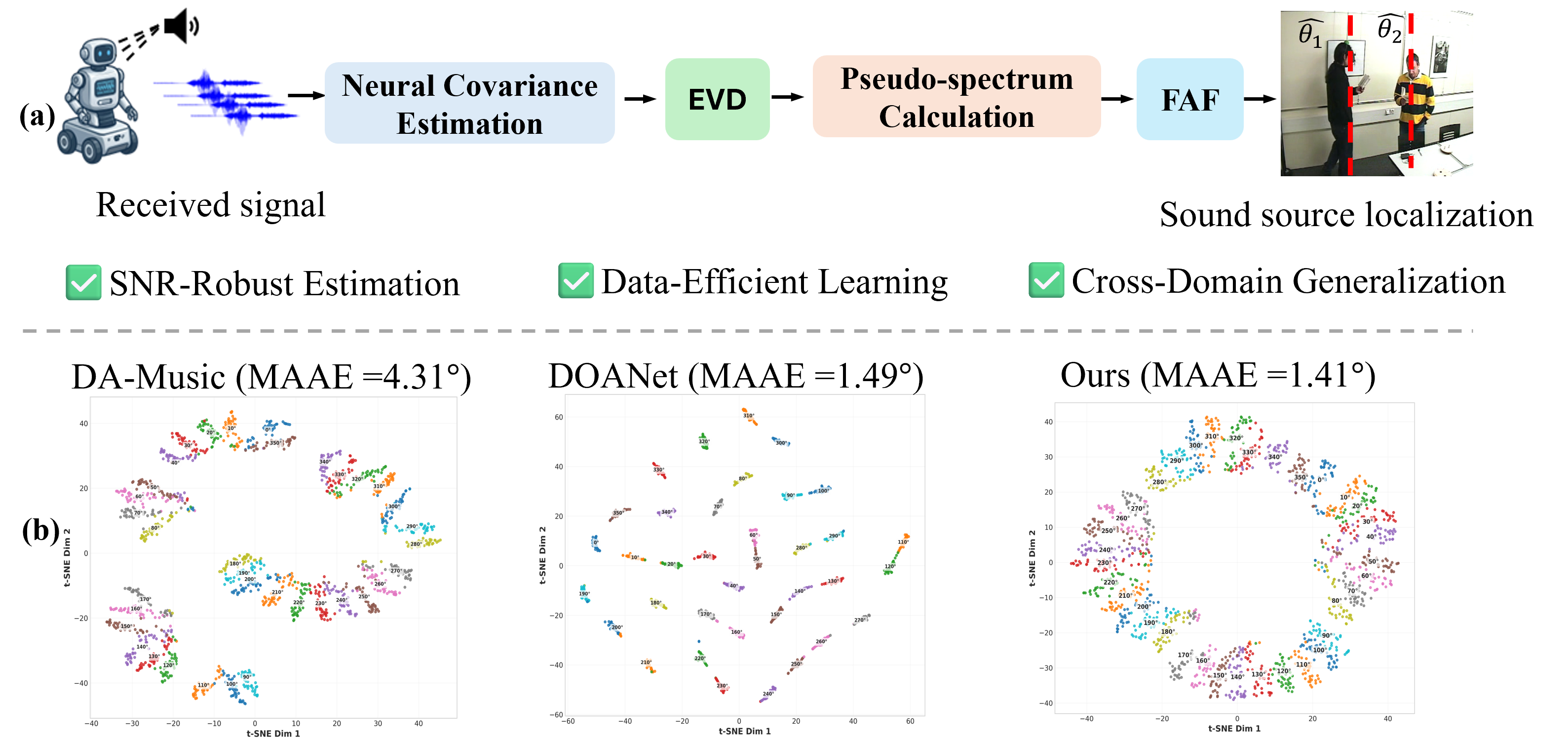}
    \caption{
    (a) Pipeline of NeuralMUSIC framework.
    (b) t-SNE visualization of DOA embeddings. The proposed method forms compact and well-separated clusters while preserving the intrinsic angular continuity of the DOA manifold, enabling physically consistent spatial representations and more reliable auditory perception for robots.
    }
    \label{intro}
    \vspace{-1mm}
\end{figure}
To overcome the limitations of both model-based and purely data-driven paradigms, hybrid SSL frameworks have recently emerged as a promising research direction. For instance, Nguyen et al. \cite{nguyen2024unet} proposed UNet-rootMUSIC, which combines a UNet with the rootMUSIC algorithm to map distorted covariance matrices to ideal ones for accurate DOA estimation under array imperfections. 
SubspaceNet \cite{shmuel2024subspacenet} employs a CNN autoencoder to learn covariance representations from empirical correlations and incorporates a differentiable Root-MUSIC layer for end-to-end training. 
Despite promising results, these hybrid methods have several limitations for robotic audition.
First, the existing approaches \cite{merkofer2023music,shmuel2024subspacenet,nguyen2024unet} are designed for narrowband radar or communication signals and do not fully address the broadband nature of acoustic signals.
Second, many methods \cite{shmuel2024subspacenet,nguyen2024unet} are restricted to specific array geometries such as uniform linear arrays (ULAs), limiting flexibility for diverse robotic platforms.
Third, introducing additional neural modules after the classical pipeline often reduces interpretability and may weaken generalization \cite{merkofer2023music}.

Another practical challenge in robotic SSL is the scarcity of labeled data.
Obtaining multichannel recordings with accurate DOA annotations is labor-intensive and often requires carefully controlled experimental setups or additional tracking systems to provide ground-truth source locations.
In contrast, large amounts of unlabeled acoustic data can be naturally collected during routine robot operation in diverse environments.
Effectively leveraging such data is therefore crucial for developing scalable and practical robot audition systems.

To address the above challenges, we propose NeuralMUSIC, a hybrid subspace framework that integrates deep representation learning with the classical MUSIC algorithm for robust DOA estimation in robotic audition. Specifically, a neural covariance encoder is designed to learn noise-resilient spatial representations while preserving the analytical structure of subspace decomposition. To address the broadband nature of acoustic signals, we introduce a Frequency Attention Fusion (FAF) module that adaptively integrates spatial cues across frequency bands. An adaptive source-number estimation module further enables reliable multi-source localization in dynamic and low-SNR environments. Additionally, a Self-supervised Spatial Correlation Learning (SSCL) strategy is developed to leverage unlabeled multichannel recordings, improving robustness while reducing reliance on annotated data. The main contributions of this work are summarized as follows:
\begin{enumerate}
\item We propose NeuralMUSIC, a unified hybrid framework that combines neural covariance learning with the classical subspace algorithm, providing accurate and reliable localization under low-SNR and reverberant environments.
\item We introduce a frequency-adaptive fusion module and a dynamic source-number estimation mechanism to enable broadband feature aggregation and robust multi-source localization.
\item We introduce a self-supervised pretraining strategy that exploits inter-channel spatial correlations from unlabeled acoustic data, improving data efficiency and accuracy.
\item Comprehensive experiments across diverse robot audition tasks validate that NeuralMUSIC achieves superior localization accuracy, robustness, and cross-environment generalization.
\end{enumerate}

\section{System Model and Preliminary}

\subsection{Signal Model}
Consider a uniform linear array (ULA) with $N$ microphones spaced by $d$, receiving $M$ far-field acoustic sources with directions of arrival (DOAs)
\begin{equation}
\boldsymbol{\theta} = [\theta_1,\ldots,\theta_M]^{\top},
\end{equation}
where $\theta_m$  $(m=1,2,...,M)$ denotes the DOA of the $m$-th source and $c$ is the speed of sound.
The signal at the $n$-th microphone is
\begin{equation}
x_n(t) = \sum_{m=1}^{M} s_m\!\left(t - \tau_n(\theta_m)\right) + v_n(t),
\end{equation}
where $s_m(t)$ is the $m$-th source signal, $v_n(t)$ is additive noise, and the propagation delay relative to the first microphone is
\[
\tau_n(\theta_m) = \frac{(n-1)d\sin\theta_m}{c}.
\]

Applying the short-time Fourier transform (STFT) gives
\begin{equation}
X_n(f,t) = \sum_{m=1}^{M} e^{-j2\pi f \tau_n(\theta_m)} S_m(f,t) + V_n(f,t),
\end{equation}
where $f$ denotes frequency and $t$ the time frame index.

Stacking all microphone signals into a vector yields the array observation model:
\begin{equation}
\mathbf{X}(f,t) = \mathbf{A}(f,\boldsymbol{\theta}) \mathbf{S}(f,t) + \mathbf{V}(f,t),
\end{equation}
where $\mathbf{X}(f,t)$, $\mathbf{S}(f,t)$, and $\mathbf{V}(f,t)$ denote the observation, source, and noise vectors. The steering matrix is
\begin{equation}
\mathbf{A}(f,\boldsymbol{\theta})
= [\mathbf{a}(f,\theta_1),\ldots,\mathbf{a}(f,\theta_M)],
\end{equation}
\begin{equation}
\mathbf{a}(f,\theta) =
[1,\; e^{-j2\pi f \tau(\theta)},\; \ldots,\; e^{-j2\pi f (N-1)\tau(\theta)}]^{\top},
\end{equation}
where $\tau(\theta)=\frac{d\sin\theta}{c}$. The formulation is presented using a ULA for clarity, but the framework is not restricted to ULA and can be extended to arbitrary microphone arrays.


\subsection{Subspace-Based Localization (MUSIC)}
For a given frequency $f$, assuming stationary and uncorrelated sources, the spatial covariance matrix of the array observation is
\begin{equation}
\begin{aligned}
\mathbf{R}_x(f) 
&= \mathbb{E}\!\left[\mathbf{X}(f,t)\mathbf{X}^{H}(f,t)\right] \\
&= \mathbf{A}(f)\mathbf{R}_s(f)\mathbf{A}^{H}(f) 
   + \sigma_v^2(f)\mathbf{I}_N.
\end{aligned}
\end{equation}
where $\mathbf{R}_s(f)=\mathbb{E}[\mathbf{S}(f)\mathbf{S}^{H}(f)]$ is the source covariance matrix, $\sigma_v^2(f)$ denotes the spatially white noise power, and $\mathbf{I}_N$ is the $N\times N$ identity matrix.

Eigenvalue decomposition (EVD) of $\mathbf{R}_x(f)$ gives
\begin{equation}
\mathbf{R}_x(f)
= \mathbf{E}_s(f)\boldsymbol{\Lambda}_s(f)\mathbf{E}_s^{H}(f)
+ \mathbf{E}_n(f)\boldsymbol{\Lambda}_n(f)\mathbf{E}_n^{H}(f),
\end{equation}
where $\mathbf{E}_s(f) \in \mathbb{C}^{N\times M}$ and $\mathbf{E}_n(f)\in\mathbb{C}^{N\times(N-M)}$ denote the signal and noise subspaces, respectively. ${\boldsymbol{\Lambda}}_s(f)$ and ${\boldsymbol{\Lambda}}_n(f)$ are diagonal matrices of the corresponding eigenvalues.

\begin{figure*}[t]
    \centering
    \includegraphics[width=0.97\linewidth]{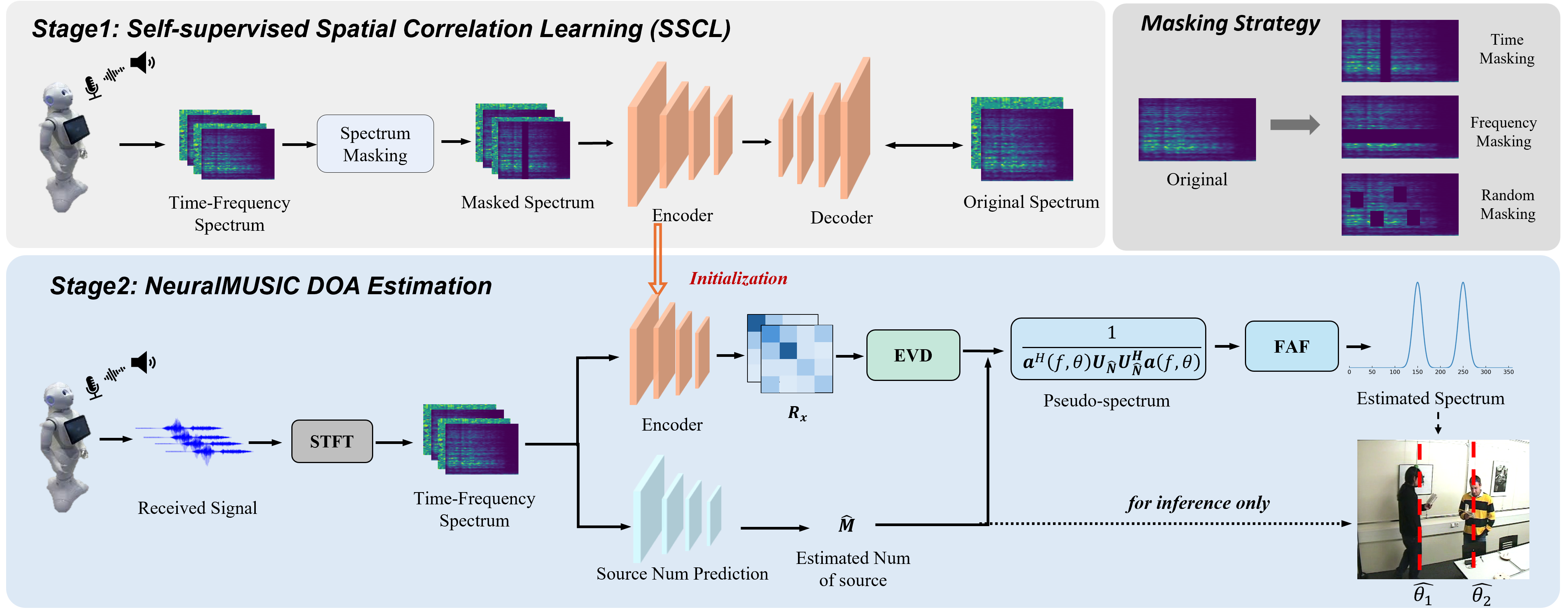}
    \caption{Overall architecture of the proposed hybrid neural–subspace framework for robot audition.}
    \label{framework}
\end{figure*}

Since the steering vectors lie in the signal subspace, they are orthogonal to the noise subspace:
\begin{equation}
\mathbf{E}_n^{H}(f)\mathbf{a}(f,\theta_m)=\mathbf{0}, \quad m=1,\ldots,M.
\end{equation}
This orthogonality leads to the MUSIC pseudo-spectrum
\begin{equation}
P_{\text{MU}}(\theta;f)
= \frac{1}{\|\mathbf{E}_n^{H}(f)\mathbf{a}(f,\theta)\|_2^2},
\end{equation}
and the DOAs are estimated from the $M$ dominant peaks of $P_{\text{MU}}(\theta;f)$. 
Since acoustic signals are inherently broadband, spanning multiple frequency bands and resulting in frequency-dependent signal and noise subspaces, many broadband MUSIC extensions have been developed to perform cross-frequency integration, commonly through non-coherent spectral averaging \cite{salvati2014incoherent} or coherent subspace focusing \cite{wang1985coherent}.

However, when deployed in real-world robotic acoustic environments, MUSIC encounters several practical challenges:
\begin{itemize}
    \item \textbf{Noise sensitivity:} Covariance estimation is easily distorted by noise and reverberation, leading to subspace leakage and degraded localization accuracy.
    \item \textbf{Source-number reliance:} The number of active sources must be specified in advance and incorrect estimation results in missed detections or spurious peaks.
    \item \textbf{Limited broadband processing:} Existing broadband extensions rely on heuristic averaging or handcrafted focusing, which reduces robustness.
\end{itemize}

These limitations motivate hybrid neural–subspace approaches that enhance covariance estimation while preserving the generalization ability of MUSIC.

\section{Proposed Method: NeuralMUSIC}
Fig.~\ref{framework} illustrates the overall architecture of the proposed method. 
The NeuralMUSIC DOA estimation framework integrates a learned spatial covariance estimator into the classical MUSIC algorithm to produce a more robust and accurate pseudo-spectrum for sound source localization, while retaining the generalization ability. 
To further improve robustness and data efficiency, SSCL is introduced to leverage unlabeled multichannel data. 
This section provides a detailed description of each module.

\subsection{Neural Covariance Matrix Estimation}
Let $\mathbf{x}(t) = [x_1(t), \ldots, x_N(t)]^{\top}$ denote the multichannel audio recorded by an $N$-microphone array. 
Applying the short-time Fourier transform (STFT) yields the complex spectrogram 
$\mathbf{X} \in \mathbb{C}^{N \times F \times L}$, where $F$ is the total number of frequency bins and $L$ is the time length.
The network input is constructed by stacking magnitude and phase of the spectrogram:
\begin{equation}
\mathbf{Y} = \big[\,|\mathbf{X}(f,t)|;\ \angle \mathbf{X}(f,t)\,\big] 
\in \mathbb{R}^{2N \times F \times L}.
\end{equation}

We first employ an encoder 
$\mathcal{F}_{\mathrm{enc}}(\cdot;\boldsymbol{\phi})$ 
to extract spatial features:
\begin{equation}
\mathbf{F_Y}
=
\mathcal{F}_{\mathrm{enc}}(\mathbf{Y};\boldsymbol{\phi})
\in
\mathbb{R}^{F\times D},
\label{eq:encoder}
\end{equation}
where $D$ denotes the latent feature dimension. 
The representation $\mathbf{F_Y}$ captures inter-channel spatial correlations across time and frequency. The encoder is implemented as a lightweight convolutional network composed of four convolutional blocks with $3\times3$ kernels and progressive max-pooling. This hierarchical design captures local time–frequency patterns in early layers and aggregates cross-microphone spatial dependencies in deeper layers. The final feature map is projected to the desired latent dimension for subsequent covariance regression.

Each frequency component in encoded feature $\mathbf{F_Y}(f)$ is then mapped to the spatial covariance matrix through a regression network:
\begin{equation}
\hat{\mathbf{R}}_x(f)
=
\mathcal{F}_{\mathrm{reg}}(\mathbf{F_Y}(f);\boldsymbol{\theta})
=
\mathbf{W}\mathbf{F_Y}(f) + \mathbf{b},
\quad
f = 1,\dots,F,
\end{equation}
where $\mathbf{W} \in \mathbb{R}^{2N^2 \times D}$ and 
$\mathbf{b} \in \mathbb{R}^{2N^2}$ are learnable parameters. 
The regressor outputs the real and imaginary components, which are reshaped and combined to form $\hat{\mathbf{R}}_x(f) \in \mathbb{C}^{N \times N}$.

To enable stable eigen-decomposition for subsequent subspace extraction, 
the estimated covariance matrix is symmetrized to enforce Hermitian structure:
\begin{equation}
\hat{\mathbf{R}}_x(f) =
\frac{1}{2}
\left(
\hat{\mathbf{R}}_x(f)
+
\hat{\mathbf{R}}_x^{H}(f)
\right).
\end{equation}
A small diagonal loading term is further added prior to eigen-decomposition to enhance numerical robustness. 
Instead of explicitly enforcing positive semi-definiteness through hard constraints, this strategy stabilizes the subspace computation while retaining flexibility for data-driven learning of spatial correlations.

\subsection{Source Number Prediction}
In real robotic environments, the number of active sound sources may vary over time. To enable adaptive subspace decomposition, we introduce a source-number prediction branch.
Given the input feature tensor $\mathbf{Y}$, a parallel neural branch 
$\mathcal{F}_{\mathrm{num}}(\cdot;\boldsymbol{\psi})$ 
is employed to predict the most likely source count:
\begin{equation}
\hat{M}
=
\mathcal{F}_{\mathrm{num}}(\mathbf{Y};\boldsymbol{\psi}),
\end{equation}
where $\hat{M} \in \{1,\dots,M_{\max}\}$ denotes the estimated number of sources. In this paper, $\mathcal{F}_{\mathrm{num}}(\cdot;\boldsymbol{\psi})$ consists of convolutional layers for extracting spatial cues from multichannel features, followed by MLPs to predict the source count. 

The predicted $\hat{M}$ determines the dimension of the signal subspace in eigenvalue decomposition, where the leading $\hat{M}$ eigenvectors are assigned to the signal subspace and the remaining correspond to the noise subspace. 
This mechanism allows the system to adapt to dynamic multi-source robotic environments without manual model-order selection.

\subsection{DOA Spectrum Estimation and Frequency Attention Fusion (FAF)}

Given the estimated covariance matrices 
$\hat{\mathbf{R}}_x(f)$ 
and the predicted source number $\hat{M}$, 
we perform eigenvalue decomposition:
\begin{equation}
\hat{\mathbf{R}}_x(f)
=
\hat{\mathbf{U}}_s(f)\hat{\boldsymbol{\Lambda}}_s(f)\hat{\mathbf{U}}_s^{H}(f)
+
\hat{\mathbf{U}}_n(f)\hat{\boldsymbol{\Lambda}}_n(f)\hat{\mathbf{U}}_n^{H}(f),
\end{equation}
where 
$\hat{\mathbf{U}}_n(f) \in \mathbb{C}^{N \times (N-\hat{M})}$ and $\hat{\mathbf{U}}_s(f) \in \mathbb{C}^{N \times \hat{M}}$ denotes the estimated noise subspace and signal subspace.  $\hat{\boldsymbol{\Lambda}}_s(f)$ and $\hat{\boldsymbol{\Lambda}}_n(f)$ are diagonal matrices of the corresponding eigenvalues.

The frequency-wise pseudo-spectrum is computed as
\begin{equation}
\hat{P}(f,\theta)
=
\frac{1}{
\mathbf{a}^{H}(f,\theta)
\hat{\mathbf{U}}_n(f)
\hat{\mathbf{U}}_n^{H}(f)
\mathbf{a}(f,\theta)
}.
\end{equation}

To aggregate information across frequency bins, we introduce a 
{Frequency Attention Fusion (FAF) module}. 
Let $\Theta = \{\theta_1,\dots,\theta_{|\Theta|}\}$
denote the predefined discrete angular grid. 
For each steering angle $\theta \in \Theta$, 
the frequency-wise spectrum is stacked as
\begin{equation}
\mathbf{p}(\theta)
=
\big[
\hat{P}(1,\theta), \ldots, \hat{P}(F,\theta)
\big]^{\top}
\in
\mathbb{R}^{F}.
\end{equation}

The attention weights are computed via global average pooling followed by a two-layer transformation:
\begin{equation}
\mathbf{z}
=
\frac{1}{|\Theta|}
\sum_{\theta \in \Theta}
\mathbf{p}(\theta)
\in
\mathbb{R}^{F},
\end{equation}
\begin{equation}
\boldsymbol{\alpha}
=
\sigma\!\left(
\mathbf{W}_{f2} \,\delta(\mathbf{W}_{f1} \mathbf{z})
\right),
\quad
\boldsymbol{\alpha} \in \mathbb{R}^{F},
\end{equation}
where $\mathbf{W}_{f1} \in \mathbb{R}^{\frac{F}{r} \times F}$,
$\mathbf{W}_{f2} \in \mathbb{R}^{F \times \frac{F}{r}}$ 
are learnable weight matrices with reduction ratio $r$,
$\delta(\cdot)$ denotes ReLU, and 
$\sigma(\cdot)$ denotes the sigmoid function. 

The final broadband spectrum is obtained by weighted aggregation:
\begin{equation}
\hat{P}(\theta)
=
\sum_{f=1}^{F}
\alpha_f \hat{P}(f,\theta),
\end{equation}
where $\alpha_f$ denotes the $f$-th element of the attention vector $\boldsymbol{\alpha}\in\mathbb{R}^{F}$. FAF adaptively reweights frequency bins, suppressing unreliable or noise-dominated components while emphasizing informative bands. This allows the model to focus on task-relevant frequency regions under different robotic tasks.

Finally, the DOA estimates $\{\hat{\theta}_m\}_{m=1}^{\hat{M}}$ are obtained by selecting the $\hat{M}$ largest local maxima of the broadband spectrum $\hat{P}(\theta)$ over the angular grid $\Theta$.


\subsection{Training Objectives}
Instead of directly regressing discrete DOA angles \cite{merkofer2023music,shmuel2024subspacenet}, the proposed network predicts a continuous broadband spectrum $\hat{P}(\theta)$ over the predefined angular grid $\Theta$. 
This representation provides smoother gradients for optimization, facilitates stable training, and naturally supports multi-source scenarios through peak detection. 
Moreover, it aligns with human auditory perception, where sound localization is often interpreted as a spatial likelihood distribution rather than a single deterministic angle.

To supervise the spectrum prediction, we construct a ground-truth spectrum $P(\theta)$ by placing Gaussian kernels centered at the true DOAs $\{\theta_i\}_{i=1}^{M}$:
\begin{equation}
P(\theta)
=
\sum_{i=1}^{M}
\exp\!\left(
-\frac{(\theta-\theta_i)^2}{2\sigma^2}
\right),
\end{equation}

The DOA loss is defined as the mean squared error:
\begin{equation}
\mathcal{L}_{\mathrm{DOA}}
=
\frac{1}{|\Theta|}
\sum_{\theta \in \Theta}
\big(
\hat{P}(\theta) - P(\theta)
\big)^2 .
\end{equation}

In addition, to supervise the source-number prediction module, we employ a cross-entropy loss:
\begin{equation}
\mathcal{L}_{\mathrm{cls}}
=
\mathrm{CE}(\hat{M}, M_{\mathrm{gt}}).
\end{equation}

The overall training objective is given by
\begin{equation}
\mathcal{L}
=
\mathcal{L}_{\mathrm{DOA}}
+
\lambda \, \mathcal{L}_{\mathrm{cls}},
\end{equation}
where $\lambda$ balances the two objectives.

\begin{table*}[t]
\centering
\vspace{8pt}
\renewcommand{\arraystretch}{1.25}
\setlength{\tabcolsep}{14pt}
\caption{MAAE ($^\circ$) comparison across different datasets and source configurations. UN denotes the unknown source-number setting. The best results are highlighted in bold, and the second-best results are underlined.}
\begin{tabular}{lcccccccc}
\toprule
\multirow{2}{*}{\textbf{Method}} 
& \multicolumn{3}{c}{\textbf{GSC}} 
& \multicolumn{3}{c}{\textbf{AV16.3}} 
& \multicolumn{2}{c}{\textbf{Single-Source Tasks}} \\
\cmidrule(lr){2-4} \cmidrule(lr){5-7} \cmidrule(lr){8-9}
& \textbf{$M$=1} & \textbf{$M$=2} & \textbf{$M$=UN}
& \textbf{$M$=1} & \textbf{$M$=2} & \textbf{$M$=UN}
& \textbf{SLoClas} & \textbf{AFPILD} \\
\midrule
MUSIC   & 3.96 & 47.77 & 25.87 & 15.45 & 33.56 & 24.51 & 8.93  & 16.02 \\
NormMUSIC    & 1.97 & 55.91 & 28.94 & 14.33 & 42.33 & 28.33 & 5.96  & 15.90 \\
Beamforming   & 2.47 & 61.47 & 31.97 & 14.39 & 32.18 & 23.29 & 7.50  & 21.49 \\
TOPS          & 5.14 & 49.78 & 27.46 & 19.26 & 25.07 & 22.17 & 16.29 & 64.40 \\
FRIDA         & 13.01 & 41.39 & 27.20 & 37.35 & 45.20 & 41.28 & 26.11 & 56.71 \\
CRNN     & 2.77 & 9.30  & 7.70  & 12.05 & 13.14 & 18.38 & \underline{3.37}  & \underline{10.55} \\
Transformer   & 1.77 & \underline{5.80}  & 4.53  & 19.26 & 13.61 & 20.24 & 3.59 & 11.16 \\
DOANet        & \underline{1.49} & 6.08  & \underline{3.90}  & 14.10 & 14.33 & 20.54 & 3.60  & 12.78 \\
DeepDAE       & 4.05 & 9.27  & 7.13  & 14.34 & 14.10 & 21.11 & 3.87  & 17.29 \\
DeepMusic     & 2.37 & 16.99 & 10.57 & \underline{10.89} & 13.40 & 16.44 & 5.80  & 14.98 \\
DA-Music      & 4.31 & 11.42 & 8.87  & 11.10 & \underline{11.45} & \underline{15.15} & 4.87  & 15.72 \\
\midrule
\textbf{Ours} 
& \textbf{1.41} & \textbf{2.25} & \textbf{2.14}
& \textbf{7.64} & \textbf{11.17} & \textbf{13.51}
& \textbf{2.91} & \textbf{10.24} \\
\bottomrule
\label{table:comparison}
\end{tabular}
\vspace{-5pt}
\end{table*}

\subsection{Self-supervised Spatial Correlation Learning (SSCL)}
To exploit unlabeled acoustic data collected during routine operation. We introduce a {Self-supervised Spatial Correlation Learning (SSCL)} strategy.

Given the input feature $\mathbf{Y}$, 
we first randomly select one microphone channel index $n^\ast \in \{1,\dots,N\}$. 
On the spectrogram of the selected channel, we apply masking operations to its magnitude and phase components, resulting in a partially observed tensor $\tilde{\mathbf{Y}}$.
Specifically, we design three masking strategies on the selected channel: 
(i) \textit{Time masking}, where contiguous time segments are set to zero. 
(ii) \textit{Frequency masking}, where continuous frequency bands are masked.
and (iii) \textit{Random masking}, where time–frequency elements are randomly dropped. 
Illustrations of these masking schemes are shown in Fig.~\ref{framework}.
Unlike conventional masked reconstruction in speech representation learning, SSCL exploits inter-channel dependencies induced by acoustic propagation, encouraging the network to capture spatial correlations.

The encoder $\mathcal{F}_{\mathrm{enc}}$ first extracts a latent representation by Eq. \ref{eq:encoder}, using the masked input $\tilde{\mathbf{Y}}$ instead of $\mathbf{Y}$. 
Then, the decoder reconstructs the masked channel:
\begin{equation}
\hat{\mathbf{Y}}_{n^\ast}
=
\mathcal{F}_{\mathrm{dec}}(\tilde{\mathbf{Y}};\boldsymbol{\eta}).
\end{equation}

The reconstruction loss is defined as
\begin{equation}
\mathcal{L}_{\mathrm{rec}}
=
\left\|
\hat{\mathbf{Y}}_{n^\ast}
-
\mathbf{Y}_{n^\ast}
\right\|_2^2 .
\end{equation}

\begin{figure}[t]
    \centering
    \includegraphics[width=0.9\linewidth]{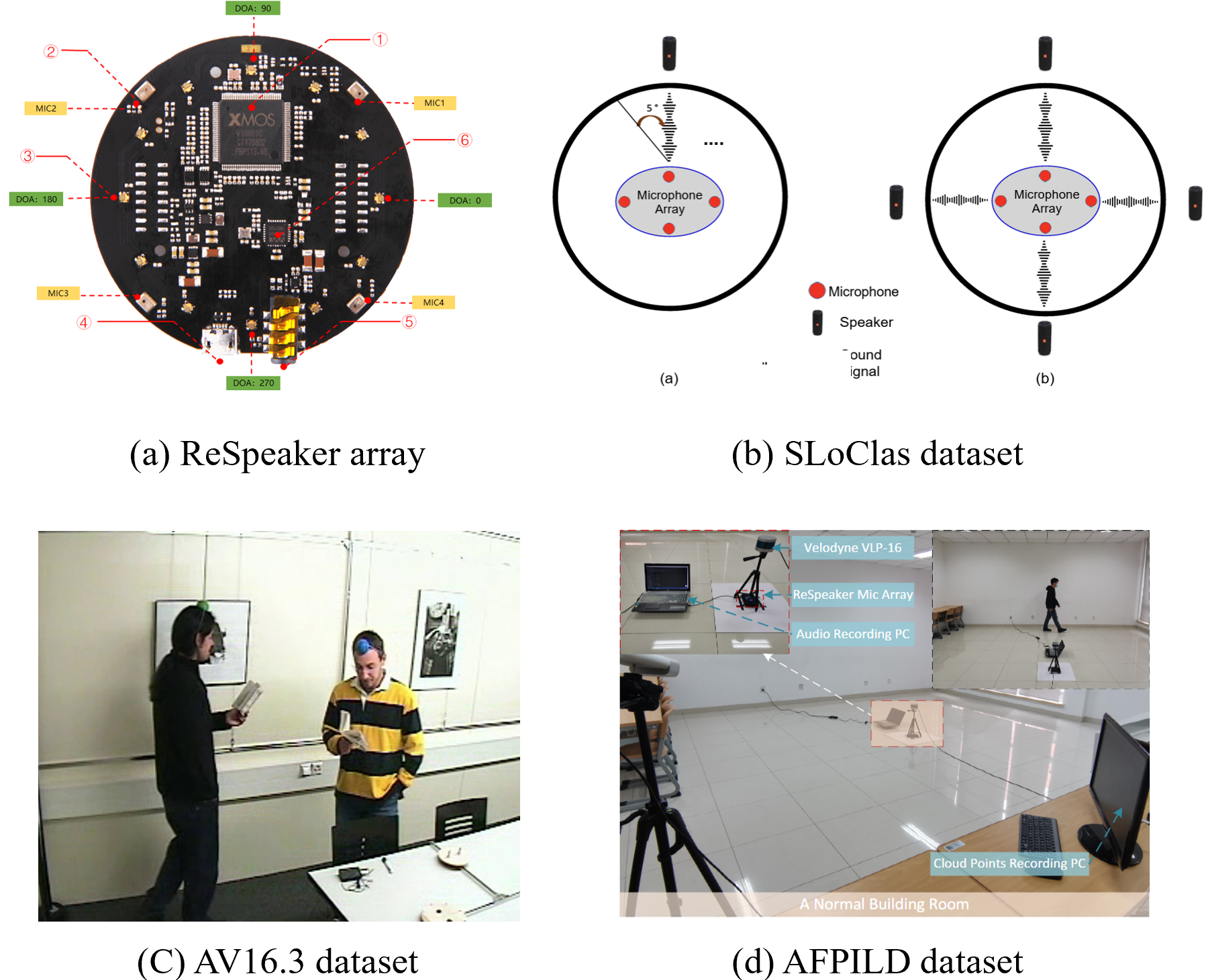}
    \caption{
    (a). ReSpeaker microphone array used in the GSC experimental setting.
    (b)–(d). Experimental environments and sample for SLoClas, AV16.3, and AFPILD datasets.}
    \label{dataset}
\end{figure}

By reconstructing the missing channel from the remaining microphones, the  encoder $\mathcal{F}_{\mathrm{enc}}$ is encouraged to capture cross-channel spatial dependencies induced by acoustic propagation delays and inter-microphone phase relationships. 
Such dependencies directly correspond to the structural information encoded in the spatial covariance matrix $\hat{\mathbf{R}}_x(f)$.
Therefore, SSCL provides a well-informed initialization for neural covariance estimation, improving robustness and data efficiency in downstream supervised training.

\begin{figure*}[t]  
    \centering
    \includegraphics[width=0.95\linewidth]{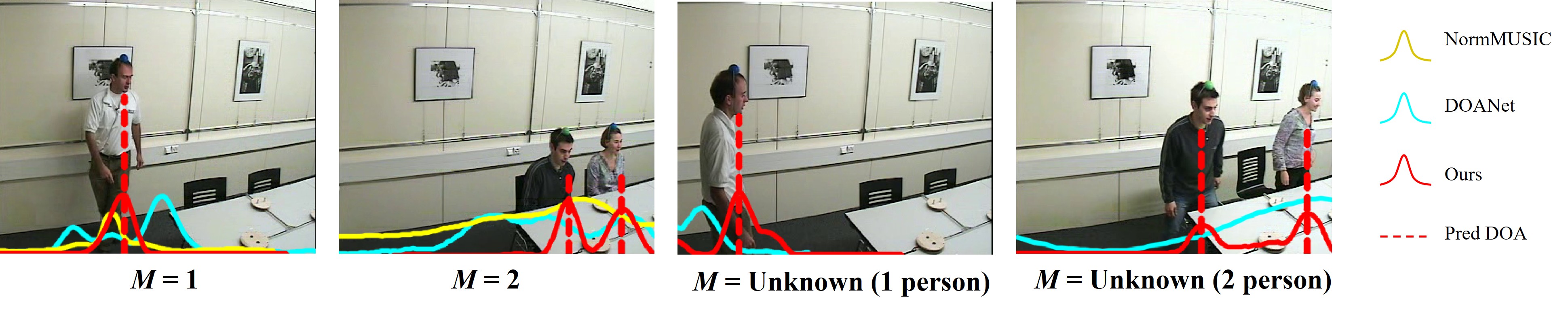}
    \caption{Qualitative comparison under different source configurations on AV16.3 dataset.}
    \label{test}
    \label{fig: test_samples}
\end{figure*}

\section{Experiments}
\subsection{Experimental Setting}
\noindent\textbf{Datasets:} We evaluate the proposed method on multiple robotic audition tasks across diverse datasets, including Google Speech Commands (GSC) \cite{warden1804speech} and AV16.3 \cite{lathoud2004av16} for speaker localization, SLoClas \cite{qian2021sloclas} for acoustic event localization, and AFPILD \cite{wu2024afpild} for pedestrian localization based on footstep sounds. 
For the GSC dataset, we select speech samples and generate multichannel recordings using the Pyroomacoustics \cite{scheibler2018pyroomacoustics} platform, incorporating room reverberation and environmental noise to emulate realistic acoustic conditions and we adopt the microphone setting as ReSpeaker\footnote{\url{https://wiki.seeedstudio.com/ReSpeaker_Mic_Array_v2.0/}} array shown in Fig. \ref{dataset} (a). More detailed experimental configurations are provided in the public repository.

\noindent\textbf{Comparison Methods and Evaluation Metrics:} 
We compare the proposed method with both classical subspace-based approaches and recent deep learning–based methods. 
The traditional baselines include MUSIC \cite{schmidt1986multiple}, NormMUSIC \cite{salvati2014incoherent}, Beamforming \cite{valin2004localization}, TOPS \cite{yoon2006tops} and FRIDA \cite{pan2017frida}. 
The learning-based methods include DeepDAE \cite{he2018deep}, DeepMusic \cite{elbir2020DeepMusic}, CRNN \cite{wu2024afpild}, Transformer \cite{kuang2022bast}, and DOANet \cite{qayyum2020DOANet}, while DA-Music \cite{merkofer2023music} is considered as a representative hybrid method that combines model-based and data-driven strategies.

To evaluate localization accuracy, we adopt the Mean Absolute Angular Error (MAAE), which accounts for angular periodicity. 
For a predicted DOA $\hat{\theta}$ and ground-truth DOA $\theta$, the circular error is computed as
\begin{equation}
e(\hat{\theta}, \theta)
=
\left|
\left( (\hat{\theta} - \theta + 180^\circ) \bmod 360^\circ \right)
- 180^\circ
\right|.
\end{equation}
The MAAE is defined as
\begin{equation}
\mathrm{MAAE}
=
\frac{1}{H}
\sum_{h=1}^{H}
e(\hat{\theta}_h, \theta_h),
\end{equation}
where $H$ denotes the number of test samples.

\begin{table}[t]
\centering
\renewcommand{\arraystretch}{1.15}
\caption{Ablation study of the proposed method on different datasets (MAAE in $^\circ$).}
\begin{tabular}{lcccc}
\toprule
\textbf{Ablation} & \textbf{GSC} & \textbf{AV16.3} & \textbf{SLoClas} & \textbf{AFPILD} \\
\midrule
w/o FAF & \underline{1.47} & \underline{8.31} & \underline{2.93} & \underline{10.42} \\
w/o SSCL  & 1.52 & 10.00 & 3.17 & 10.69 \\
Ours           & \textbf{1.41} & \textbf{7.64} & \textbf{2.91} & \textbf{10.24} \\
\bottomrule
\label{tab:ablation}
\end{tabular}
\end{table}

\noindent\textbf{Implementation Details:}
In the experiments, the STFT is computed using a window length and FFT size of 512, with a hop length of 256. 
The resulting spectrograms are resized to $257 \times 64$ for all datasets. The azimuth search space $\Theta$ is discretized over $[0^\circ, 360^\circ)$ with a resolution of $1^\circ$.
The network is trained using the Adam optimizer with an initial learning rate of $1 \times 10^{-3}$ for 100 epochs, and the $\lambda$ is set to 0.5. The batch size is set to 32. 
All deep-learning models are trained on a NVIDIA RTX 4090 GPU. 
In the experiments, we focus on azimuth estimation, as it aligns with typical planar microphone configurations in robotic audition and provides the primary cue for horizontal localization. For classical methods requiring the source number a priori, the ground-truth source count is provided to avoid unfair penalization in the unknown-source setting.

\subsection{Comparison Experiments}
The performance of different methods is shown in Table~\ref{table:comparison}. 
On GSC dataset, the proposed method consistently achieves the lowest MAAE across all source configurations.
Notably, under multi-source conditions our method significantly outperforms both traditional approaches and deep learning baselines, reducing the error from 5.80$^\circ$ (Transformer) and 6.08$^\circ$ (DOANet) to 2.25$^\circ$. 
On the real-world AV16.3 dataset, our method also achieves the best performance in all configurations. For single speaker localization, the proposed method achieves a MAAE of 7.64 $^\circ$, which is 6.69 $^\circ$ and 3.25$^\circ$ lower than NormMUSIC and DeepMusic.
In the more challenging unknown source number setting, it achieves the MAAE of 13.51$^\circ$, outperforming all other compared methods.
The results demonstrate the effectiveness of the proposed framework under real acoustic conditions with reverberation and environmental noise. Some of the localization results are visualized in Fig. \ref{fig: test_samples}. The proposed method accurately predicts the number of active sound sources under different scenarios. Meanwhile, it generates precise pseudo-spectra with sharp peaks aligned with the true DOAs.

For acoustic event localization (SLoClas) and footstep-based pedestrian localization (AFPILD) task, the proposed method also achieves the lowest errors of 2.91$^\circ$ and 10.24$^\circ$, respectively.
Compared with the best competing deep model (Transformer), our method improves the localization accuracy by 1.32$^\circ$ and 0.92$^\circ$.
Overall, the proposed method demonstrates consistent superiority across diverse tasks, source configurations, and acoustic conditions.
%

\begin{figure}[t]
    \centering

    \includegraphics[width=0.85\linewidth]{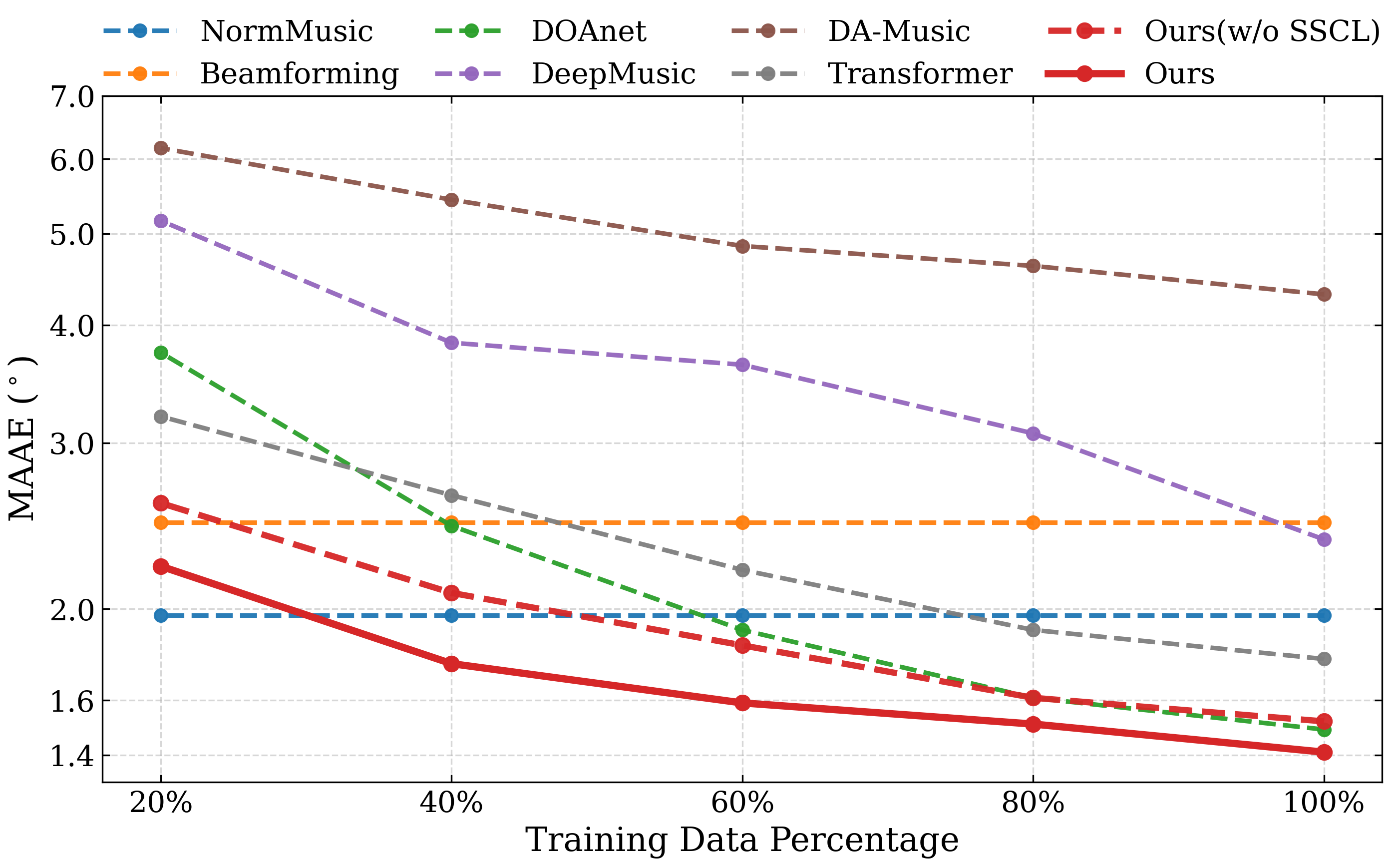}
    \caption{Performance versus different training sample ratios.}
    \label{fig:samples_results}
\end{figure}

\subsection{Ablation Study}
To investigate the contribution of each proposed component, we conduct ablation experiments by removing the Frequency Attention Fusion (FAF) module and the self-supervised Spatial Correlation Learning (SSCL) strategy, respectively. 
Table~\ref{tab:ablation} reports the MAAE results on all evaluated datasets.
Removing the FAF module leads to consistent performance degradation on most datasets. 
On AV16.3, the MAAE increases from 7.64$^\circ$ to 8.31$^\circ$, and similar degradations are observed on other datasets. 
This confirms that frequency-wise adaptive weighting improves broadband spectrum aggregation by suppressing noisy or less informative frequency components while emphasizing reliable spectral cues. Such adaptive reweighting enhances the stability of DOA estimation under varying acoustic conditions.
When SSCL is removed, the performance drops more noticeably.
This demonstrates that learning inter-channel spatial correlations from unlabeled acoustic data provides a better initialization for covariance estimation and enhances robustness in practical robotic tasks. And the full model consistently achieves the best performance across all datasets. 

\begin{figure}[t]
    \centering
    \includegraphics[width=0.9\linewidth]{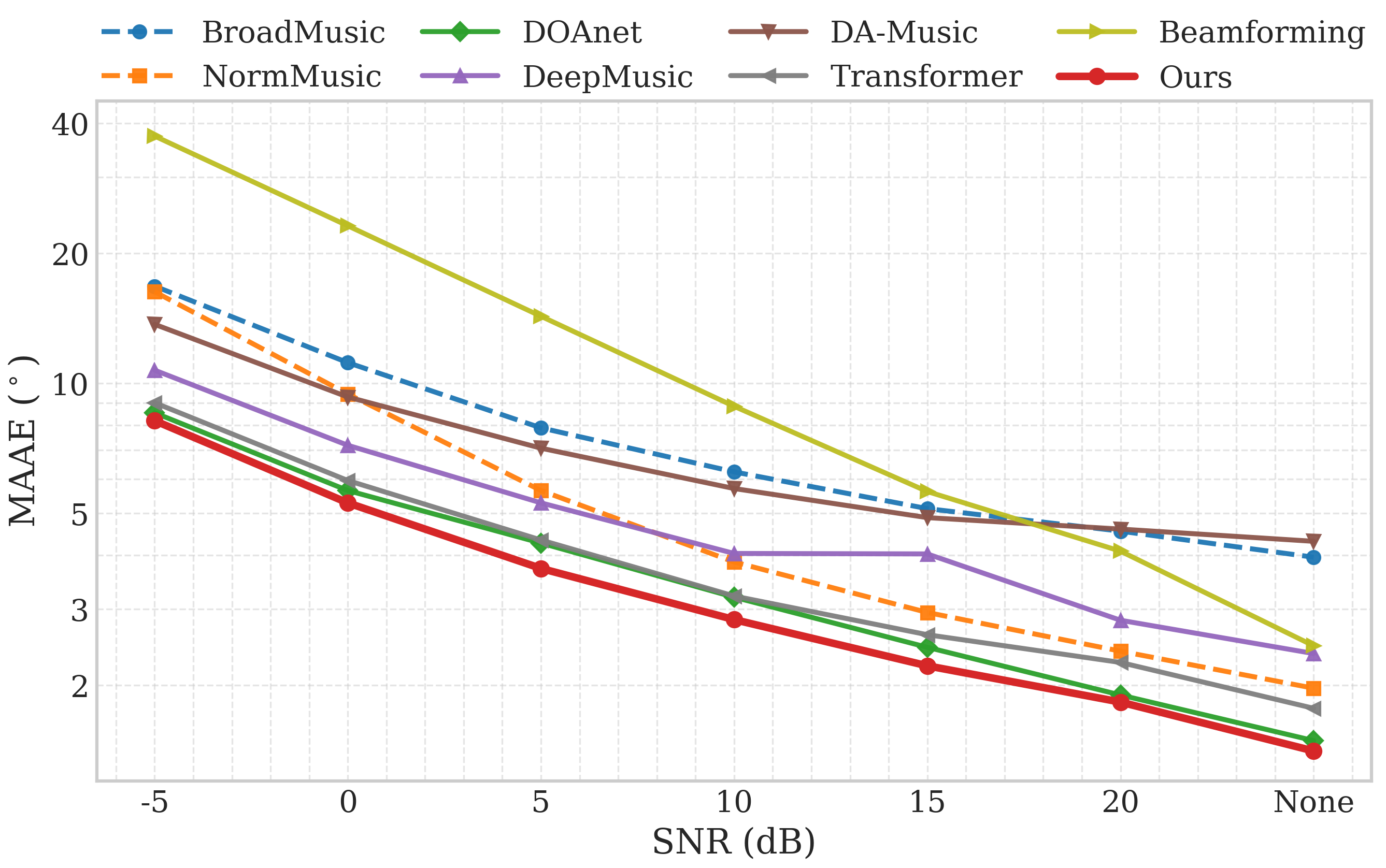}
    \caption{Performance versus different SNR.}
    \label{results_noise}
\end{figure}

\subsection{Performance under Different Training Data Ratios}
To evaluate data efficiency, we train the models using different proportions of the GSC training set under the single-source setting. The results are shown in Fig.~\ref{fig:samples_results}, where representative strong baselines are included for comparison.
As the training data decreases, all learning-based methods exhibit performance degradation. 
However, even without SSCL, our base model remains more data-efficient than other baselines. 
This improved data efficiency may stem from the proposed neural–subspace design. Instead of directly regressing angles, the network learns a structured pseudo-spectrum guided by spatial covariance modeling and subspace decomposition, which imposes strong physical priors and stabilizes training under limited supervision.

Further incorporating SSCL brings additional improvements. 
At 40\% training data, the error decreases from 1.83$^\circ$ to 1.59$^\circ$ after introducing SSCL. 
The gain becomes more evident as the amount of labeled data reduces, demonstrating that SSCL enhances spatial representation learning.
Overall, the proposed framework is inherently data-efficient, and SSCL further enhances this advantage, making the approach well suited for robotic tasks where collecting large-scale annotated acoustic data is costly and time-consuming.

\subsection{Robustness under Varying SNR Conditions}
We evaluate robustness under SNR levels from $-5$ dB to $20$ dB on GSC dataset, as shown in Fig. \ref{results_noise}. As SNR decreases, our method consistently maintains competitive or superior performance across all noise levels.
Specifically, at low SNR of $0$ dB, most methods exhibit significant degradation, while the proposed method achieves $5.29^\circ$, outperforming other approaches. 
Notably, our method consistently outperforms traditional subspace-based approaches even under high SNR conditions. 
This gain is attributed to the learned covariance representation and adaptive frequency fusion, which stabilize spatial estimation while highlighting informative frequency bands.

\begin{figure}[t]
    \centering
    \includegraphics[width=0.9\linewidth]{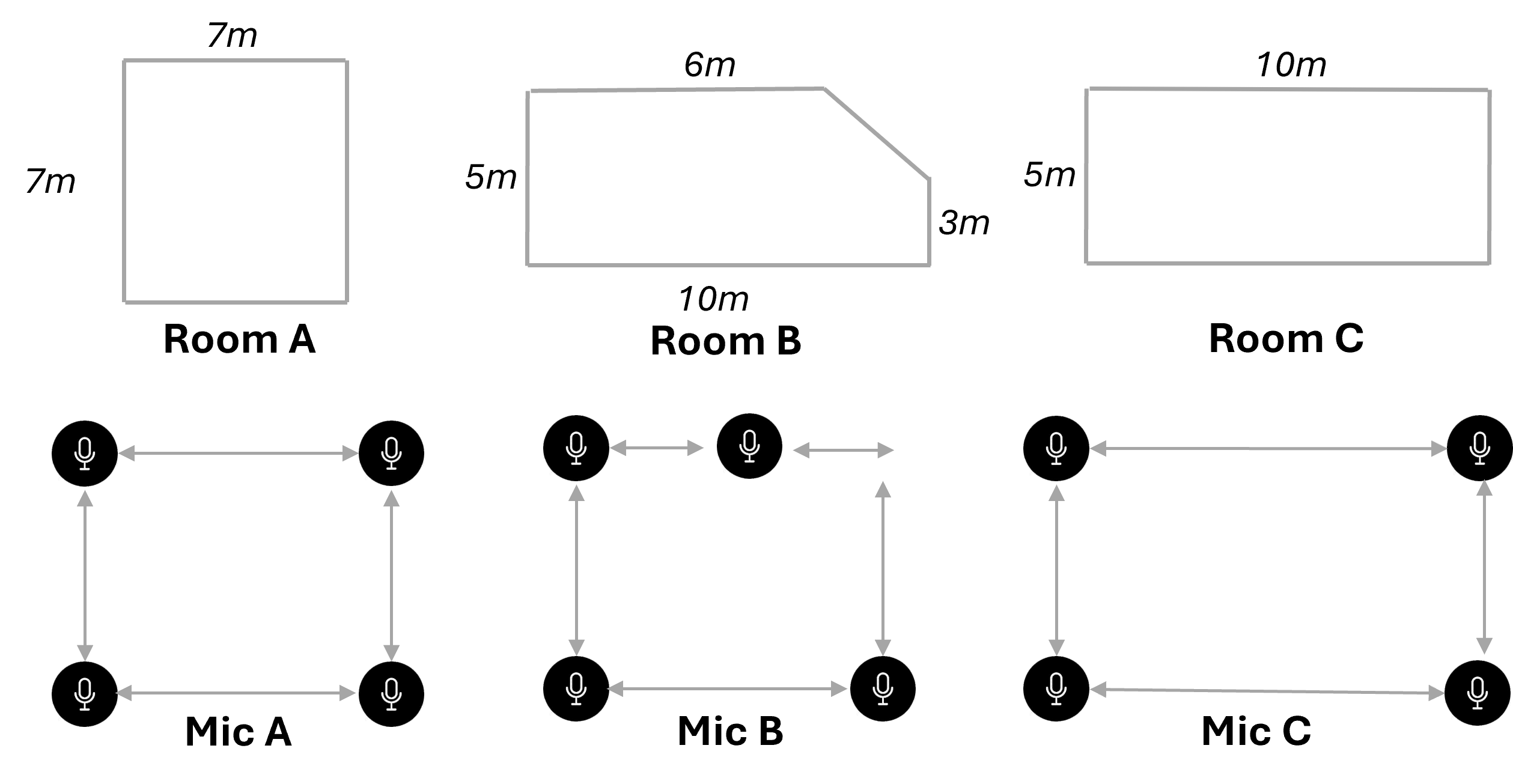}
    \caption{Illustration of the three rooms and microphone array configurations used for generalization evaluation.}
    \label{generalization_test}
\end{figure}
\subsection{Generalization Testing}

\begin{table}[t]
\centering
\renewcommand{\arraystretch}{1.1}
\setlength{\tabcolsep}{10pt}
\caption{Comparison of MAAE ($^{\circ}$) under cross-room transfer (Room A$\rightarrow$Room B and Room A$\rightarrow$Room C) with and without one-shot fine-tuning.}
\begin{tabular}{lcccc}
\toprule
\multirow{2}{*}{\textbf{Method}} & \multicolumn{2}{c}{\textbf{Room A$\rightarrow$Room B}} & \multicolumn{2}{c}{\textbf{Room A$\rightarrow$Room C}} \\
 & \textbf{w/o FT} & \textbf{w/ FT} & \textbf{w/o FT} & \textbf{w/ FT} \\
\midrule
Music     & 5.40 & 5.40 & 4.69 & 4.69 \\
NormMUSIC      & 3.27 & 3.27 & 2.45 & 2.45 \\
Beamforming    & 3.37 & 3.37 & 3.56 & 3.56 \\
TOPS           & 6.81 & 6.81 & 5.19 & 5.19 \\
FRIDA & 14.48 & 14.48 & 12.02 & 12.02 \\
CRNN   &4.36	&3.87	&3.38	&3.31 \\
Transformer & 3.51	&2.61	&2.47	&2.30 \\
DOANet         & \underline{2.87} & \underline{2.29} & \underline{2.18} & \underline{2.07}  \\
DeepDAE        & 6.12 & 6.34 & 4.70 & 5.96 \\
DeepMusic      & 4.38 & 2.77 & 3.56 & 2.56 \\
DA-Music       & 5.41 & 4.48 & 4.78 & 3.99 \\
\hline
\textbf{Ours}  & \textbf{2.20} & \textbf{1.85} & \textbf{1.71} & \textbf{1.65} \\
\bottomrule
\end{tabular}
\label{tab:cross_room}
\end{table}

To evaluate generalization, we further conduct transfer experiments.
Two scenarios are considered as shown in Fig. \ref{generalization_test}: 
(i) cross-array transfer, where the microphone geometry is changed, and 
(ii) cross-room transfer, where the acoustic environment differs. 
For each scenario, we evaluate both direct transfer (w/o FT) and one-shot fine-tuning (w/ FT), 
where only one labeled sample per discretized direction in $\Theta$ is used for adaptation.

\textbf{Cross-Room Transfer.}
Table~\ref{tab:cross_room} shows similar trends under room changes. 
Our method consistently achieves the lowest MAAE both without and with fine-tuning, demonstrating strong robustness to environmental variations.
\textbf{Cross-Array Transfer.}
As shown in Table~\ref{tab:cross_array}, most learning-based methods degrade under geometry mismatch. In contrast, our method maintains competitive accuracy without fine-tuning (2.42$^\circ$ and 3.07$^\circ$), and further improves to 1.74$^\circ$ and 1.83$^\circ$ with one-shot adaptation, achieving the best performance. However, compared with classical methods, the proposed approach shows slightly weaker generalization under large array mismatches, likely due to its data-driven covariance regression being influenced by array-specific characteristics. 

Overall, the proposed framework exhibits superior transferability across array geometries and acoustic environments compared with existing deep-learning-based methods.
This property is essential for real-world deployment, where microphone configurations may change and robots are required to operate in unseen environments without extensive retraining. 


\noindent\textbf{Feature Representation Analysis}
To analyze the learned spatial representations, we visualize the latent features using t-SNE in Fig.~\ref{intro}(b). 
Compared with baseline methods that produce fragmented clusters, NeuralMUSIC forms a smooth and continuous manifold, where neighboring angular directions are arranged sequentially, revealing a clear circular topology consistent with azimuth periodicity. 
This structured embedding preserves angular continuity and contributes to more stable localization and improved generalization.


\begin{table}[t]
\centering
\renewcommand{\arraystretch}{1.1}
\setlength{\tabcolsep}{10pt}
\caption{Comparison of MAAE ($^{\circ}$) under cross-array transfer (Mic A$\rightarrow$Mic B and Mic A$\rightarrow$Mic C) with and without one-shot fine-tuning.}
\begin{tabular}{lcccc}
\toprule
\multirow{2}{*}{\textbf{Methods}} & \multicolumn{2}{c}{\textbf{Mic A$\rightarrow$Mic B}} & \multicolumn{2}{c}{\textbf{Mic A$\rightarrow$Mic C}} \\
 & \textbf{w/o FT} & \textbf{w/ FT} & \textbf{w/o FT} & \textbf{w/ FT} \\
\midrule
Music     & 4.41 & 4.41 & 5.19 & 5.19 \\
NormMUSIC      & \textbf{2.41} & 2.41 & \underline{2.91} & 2.91 \\
TOPS            & 5.78 & 5.78 & 6.39 & 6.39 \\
Beamforming     & 2.73 & 2.73 & \textbf{2.83} & 2.83 \\
FRIDA & 15.54 & 15.54 & 16.38 & 16.38 \\
CRNN &5.64	&3.73	&6.42	&3.56 \\
Transformer &5.00	&2.33	&5.54	&2.27 \\
DOANet          & 5.12 & \underline{1.98} & 5.94 & \underline{2.02} \\
DeepDAE         & 6.52 & 5.26 & 6.91 & 5.48 \\
DeepMusic       & 7.31 & 2.96 & 5.81 & 3.01 \\
DA-Music        & 7.89 & 4.48 & 7.88 & 5.10 \\
\hline
\textbf{Ours}   & \underline{2.42} & \textbf{1.74} & 3.07 & \textbf{1.83} \\
\bottomrule
\end{tabular}
\label{tab:cross_array}
\end{table}

\section{Conclusion}\label{sec:conclu}
This paper presents a hybrid neural–subspace framework for robust sound source localization in robotic audition. 
The proposed method learns a spatial covariance representation from time–frequency features and integrates it into a MUSIC pipeline with adaptive frequency fusion, combining the representation capacity of neural networks with the generalization ability of subspace methods.  
A self-supervised spatial correlation learning strategy is further introduced to improve data efficiency.
Extensive experiments on various robotic datasets demonstrate superior accuracy, noise robustness, and strong generalization across different environments. Future work will focus on geometry-invariant learning to strengthen cross-configuration generalization. In addition, we aim to further improve data efficiency to enable reliable localization under few-shot or even one-shot training settings.

\bibliographystyle{ieeetr}        
\bibliography{main} 

@article{schmidt1986multiple,
  title={Multiple emitter location and signal parameter estimation},
  author={Schmidt, Ralph},
  journal={IEEE transactions on antennas and propagation},
  volume={34},
  number={3},
  pages={276--280},
  year={1986},
  publisher={IEEE}
}

@article{wang1985coherent,
  title={Coherent signal-subspace processing for the detection and estimation of angles of arrival of multiple wide-band sources},
  author={Wang, Hong and Kaveh, Mostafa},
  journal={IEEE Transactions on Acoustics, Speech, and Signal Processing},
  volume={33},
  number={4},
  pages={823--831},
  year={1985},
  publisher={IEEE}
}

@inproceedings{lathoud2004av16,
  title={AV16. 3: An audio-visual corpus for speaker localization and tracking},
  author={Lathoud, Guillaume and Odobez, Jean-Marc and Gatica-Perez, Daniel},
  booktitle={International Workshop on Machine Learning for Multimodal Interaction},
  pages={182--195},
  year={2004},
  organization={Springer}
}

@article{yoon2006tops,
  title={TOPS: New DOA estimator for wideband signals},
  author={Yoon, Yeo-Sun and Kaplan, Lance M and McClellan, James H},
  journal={IEEE Transactions on Signal processing},
  volume={54},
  number={6},
  pages={1977--1989},
  year={2006},
  publisher={IEEE}
}

@article{elbir2020deepmusic,
  title={DeepMUSIC: Multiple signal classification via deep learning},
  author={Elbir, Ahmet M},
  journal={IEEE Sensors Letters},
  volume={4},
  number={4},
  pages={1--4},
  year={2020},
  publisher={IEEE}
}

@article{merkofer2023music,
  title={DA-MUSIC: Data-driven DoA estimation via deep augmented MUSIC algorithm},
  author={Merkofer, Julian P and Revach, Guy and Shlezinger, Nir and Routtenberg, Tirza and Van Sloun, Ruud JG},
  journal={IEEE Transactions on Vehicular Technology},
  volume={73},
  number={2},
  pages={2771--2785},
  year={2023},
  publisher={IEEE}
}

@article{wu2024afpild,
  title={AFPILD: Acoustic footstep dataset collected using one microphone array and LiDAR sensor for person identification and localization},
  author={Wu, Shichao and Huang, Shouwang and Liu, Zicheng and Zhang, Qianyi and Liu, Jingtai},
  journal={Information Fusion},
  volume={104},
  pages={102181},
  year={2024},
  publisher={Elsevier}
}

@article{shmuel2024subspacenet,
  title={SubspaceNet: Deep learning-aided subspace methods for DoA estimation},
  author={Shmuel, Dor H and Merkofer, Julian P and Revach, Guy and Van Sloun, Ruud JG and Shlezinger, Nir},
  journal={IEEE Transactions on Vehicular Technology},
  year={2024},
  publisher={IEEE}
}

@article{qayyum2020doanet,
  title={DOANet: a deep dilated convolutional neural network approach for search and rescue with drone-embedded sound source localization},
  author={Qayyum, Alif Bin Abdul and Hassan, KM Naimul and Anika, Adrita and Shadiq, Md Farhan and Rahman, Md Mushfiqur and Islam, Md Tariqul and Imran, Sheikh Asif and Hossain, Shahruk and Haque, Mohammad Ariful},
  journal={EURASIP Journal on Audio, Speech, and Music Processing},
  volume={2020},
  number={1},
  pages={16},
  year={2020},
  publisher={Springer}
}

@inproceedings{ji2024transmusic,
  title={TransMUSIC: A transformer-aided subspace method for DOA estimation with low-resolution ADCs},
  author={Ji, Junkai and Mao, Wei and Xi, Feng and Chen, Shengyao},
  booktitle={ICASSP 2024-2024 IEEE International Conference on Acoustics, Speech and Signal Processing (ICASSP)},
  pages={8576--8580},
  year={2024},
  organization={IEEE}
}

@article{nguyen2024unet,
  title={UNet-rootMUSIC: A high accuracy direction of arrival estimation method under array imperfection},
  author={Nguyen, Duy-Thai and Le, Thanh-Hai and Doan, Van-Sang and Hoang, Van-Phuc},
  journal={AEU-International Journal of Electronics and Communications},
  volume={173},
  pages={155008},
  year={2024},
  publisher={Elsevier}
}

@inproceedings{yang2023av,
  title={Av-pedaware: Self-supervised audio-visual fusion for dynamic pedestrian awareness},
  author={Yang, Yizhuo and Yuan, Shenghai and Cao, Muqing and Yang, Jianfei and Xie, Lihua},
  booktitle={2023 IEEE/RSJ International Conference on Intelligent Robots and Systems (IROS)},
  pages={1871--1877},
  year={2023},
  organization={IEEE}
}

@inproceedings{masuyama2020self,
  title={Self-supervised neural audio-visual sound source localization via probabilistic spatial modeling},
  author={Masuyama, Yoshiki and Bando, Yoshiaki and Yatabe, Kohei and Sasaki, Yoko and Onishi, Masaki and Oikawa, Yasuhiro},
  booktitle={2020 IEEE/RSJ International Conference on Intelligent Robots and Systems (IROS)},
  pages={4848--4854},
  year={2020},
  organization={IEEE}
}

@article{schulz2021hearing,
  title={Hearing what you cannot see: Acoustic vehicle detection around corners},
  author={Schulz, Yannick and Mattar, Avinash Kini and Hehn, Thomas M and Kooij, Julian FP},
  journal={IEEE Robotics and Automation Letters},
  volume={6},
  number={2},
  pages={2587--2594},
  year={2021},
  publisher={IEEE}
}

@inproceedings{liu2025sound,
  title={Sound Source Localization for Human-Robot Interaction in Outdoor Environments},
  author={Liu, Victor and Du, Timothy and Sehn, Jordy and Collier, Jack and Grondin, Fran{\c{c}}ois},
  booktitle={2025 IEEE/RSJ International Conference on Intelligent Robots and Systems (IROS)},
  pages={6121--6126},
  year={2025},
  organization={IEEE}
}

@article{salvati2014incoherent,
  title={Incoherent frequency fusion for broadband steered response power algorithms in noisy environments},
  author={Salvati, Daniele and Drioli, Carlo and Foresti, Gian Luca},
  journal={IEEE Signal Processing Letters},
  volume={21},
  number={5},
  pages={581--585},
  year={2014},
  publisher={IEEE}
}

@article{warden1804speech,
  title={Speech commands: A dataset for limited-vocabulary speech recognition. arXiv 2018},
  author={Warden, Pete},
  journal={arXiv preprint arXiv:1804.03209},
  year={1804}
}

@inproceedings{qian2021sloclas,
  title={Sloclas: A database for joint sound localization and classification},
  author={Qian, Xinyuan and Sharma, Bidisha and El Abridi, Amine and Li, Haizhou},
  booktitle={2021 24th Conference of the Oriental COCOSDA International Committee for the Co-ordination and Standardisation of Speech Databases and Assessment Techniques (O-COCOSDA)},
  pages={128--133},
  year={2021},
  organization={IEEE}
}

@inproceedings{scheibler2018pyroomacoustics,
  title={Pyroomacoustics: A python package for audio room simulation and array processing algorithms},
  author={Scheibler, Robin and Bezzam, Eric and Dokmani{\'c}, Ivan},
  booktitle={2018 IEEE international conference on acoustics, speech and signal processing (ICASSP)},
  pages={351--355},
  year={2018},
  organization={IEEE}
}

@inproceedings{he2018deep,
  title={Deep neural networks for multiple speaker detection and localization},
  author={He, Weipeng and Motlicek, Petr and Odobez, Jean-Marc},
  booktitle={2018 IEEE International Conference on Robotics and Automation (ICRA)},
  pages={74--79},
  year={2018},
  organization={IEEE}
}

@inproceedings{valin2004localization,
  title={Localization of simultaneous moving sound sources for mobile robot using a frequency-domain steered beamformer approach},
  author={Valin, J-M and Michaud, Fran{\c{c}}ois and Hadjou, Brahim and Rouat, Jean},
  booktitle={IEEE International Conference on Robotics and Automation, 2004. Proceedings. ICRA'04. 2004},
  volume={1},
  pages={1033--1038},
  year={2004},
  organization={IEEE}
}

@article{kuang2022bast,
  title={BAST: Binaural audio spectrogram transformer for binaural sound localization},
  author={Kuang, Sheng and Shi, Jie and van der Heijden, Kiki and Mehrkanoon, Siamak},
  journal={arXiv preprint arXiv:2207.03927},
  year={2022}
}

@inproceedings{yang2024kidnappable,
  title={The un-kidnappable robot: Acoustic localization of sneaking people},
  author={Yang, Mengyu and Grady, Patrick and Brahmbhatt, Samarth and Vasudevan, Arun Balajee and Kemp, Charles C and Hays, James},
  booktitle={2024 IEEE International Conference on Robotics and Automation (ICRA)},
  pages={985--992},
  year={2024},
  organization={IEEE}
}

@inproceedings{liu2010continuous,
  title={Continuous sound source localization based on microphone array for mobile robots},
  author={Liu, Hong and Shen, Miao},
  booktitle={2010 IEEE/RSJ International Conference on Intelligent Robots and Systems},
  pages={4332--4339},
  year={2010},
  organization={IEEE}
}

@inproceedings{pan2017frida,
  title={FRIDA: FRI-based DOA estimation for arbitrary array layouts},
  author={Pan, Hanjie and Scheibler, Robin and Bezzam, Eric and Dokmani{\'c}, Ivan and Vetterli, Martin},
  booktitle={2017 IEEE International Conference on Acoustics, Speech and Signal Processing (ICASSP)},
  pages={3186--3190},
  year={2017},
  organization={IEEE}
}

@ARTICLE{8656587,
  author={Qian, Xinyuan and Brutti, Alessio and Lanz, Oswald and Omologo, Maurizio and Cavallaro, Andrea},
  journal={IEEE Transactions on Multimedia}, 
  title={Multi-Speaker Tracking From an Audio–Visual Sensing Device}, 
  year={2019},
  volume={21},
  number={10},
  pages={2576-2588},
  doi={10.1109/TMM.2019.2902489}}

\appendix
\subsection{Datasets and Experimental Setup Details}
In the experiments, we compare the performance of the on four datasets, Google Speech Commands (GSC) \cite{warden1804speech}, SoClas \cite{qian2021sloclas}, AFPILD \cite{wu2024afpild} and AV16.3 dataset \cite{lathoud2004av16}. These datasets cover different common robotics tasks including speaker localization, pedestrian detection and acoustic event localization. This section gives a detail description on the setting of each dataset.

\textbf{GSC Dataset}.
To train the proposed model, we construct a simulated acoustic dataset using speech signals from the Google Speech Commands (GSC) dataset as sound sources. Room acoustics are simulated using the Pyroomacoustics \cite{scheibler2018pyroomacoustics} toolkit with the image-source method. 
The simulated environment consists of a $7 \times 7 \times 3$ m room with a sampling rate of 16 kHz, and additive noise is added with a signal-to-noise ratio (SNR) of 30 dB. 
The sound source is randomly positioned around the microphone array with a distance ranging from 0.5 m to 2.0 m and an azimuth angle uniformly distributed between $0^\circ$ and $359^\circ$ with a resolution of $1^\circ$. 
For each azimuth angle, 100 samples are generated for training and 25 samples are generated for testing, resulting in a uniformly distributed angular dataset for supervised direction-of-arrival (DOA) estimation.

\textbf{AV16.3 Dataset}.
AV16.3 is a widely used benchmark dataset for speaker localization and tracking in indoor environments. 
Following the AV3T \cite{8656587} preprocessing pipeline, the raw recordings are processed and the dataset is divided such that sequences \textit{seq8} and \textit{seq18} are used for testing, while the remaining sequences are used for training. 
This dataset is used to evaluate the generalization ability of the proposed method under multi-speaker conditions.

\textbf{SoClas Dataset \cite{qian2021sloclas}}.
The SoClas dataset contains acoustic recordings of various sound events designed for joint sound localization and classification. 
In our setup, the microphone array is placed at the center of the room while the sound source is positioned at a fixed distance of 1.5 m. 
The azimuth angle ranges from $1^\circ$ to $360^\circ$ with a $5^\circ$ interval. 
For each direction, 80\% of the samples are randomly selected for training and the remaining 20\% are used for testing in the experiments. 
This dataset evaluates the robustness of the proposed method for event-level sound localization.

\textbf{AFPILD Dataset}.
The AFPILD dataset focuses on pedestrian localization using footstep sounds recorded by a microphone array and a LiDAR sensor. 
We follow the official training and testing split provided by the dataset authors and conduct experiments using the cloth covariant configuration ($\texttt{covariant\_type}=\texttt{'cloth'}$) according to the standard evaluation protocol. 
This dataset reflects a realistic robotic perception scenario for surrounding pedestrian localization. 

It is worth noting that the SoClas and AFPILD datasets do not explicitly provide the world coordinate system, microphone coordinate system, or corresponding calibration files. As a result, traditional DOA estimation methods may exhibit systematic angular offsets when applied directly to these datasets. To ensure a fair comparison, we estimate the optimal angular offset between the predicted DOA and the ground truth during evaluation and report the results using the offset that yields the best performance for traditional methods.

\begin{figure}[t]
    \centering
    \includegraphics[width=0.95\linewidth]{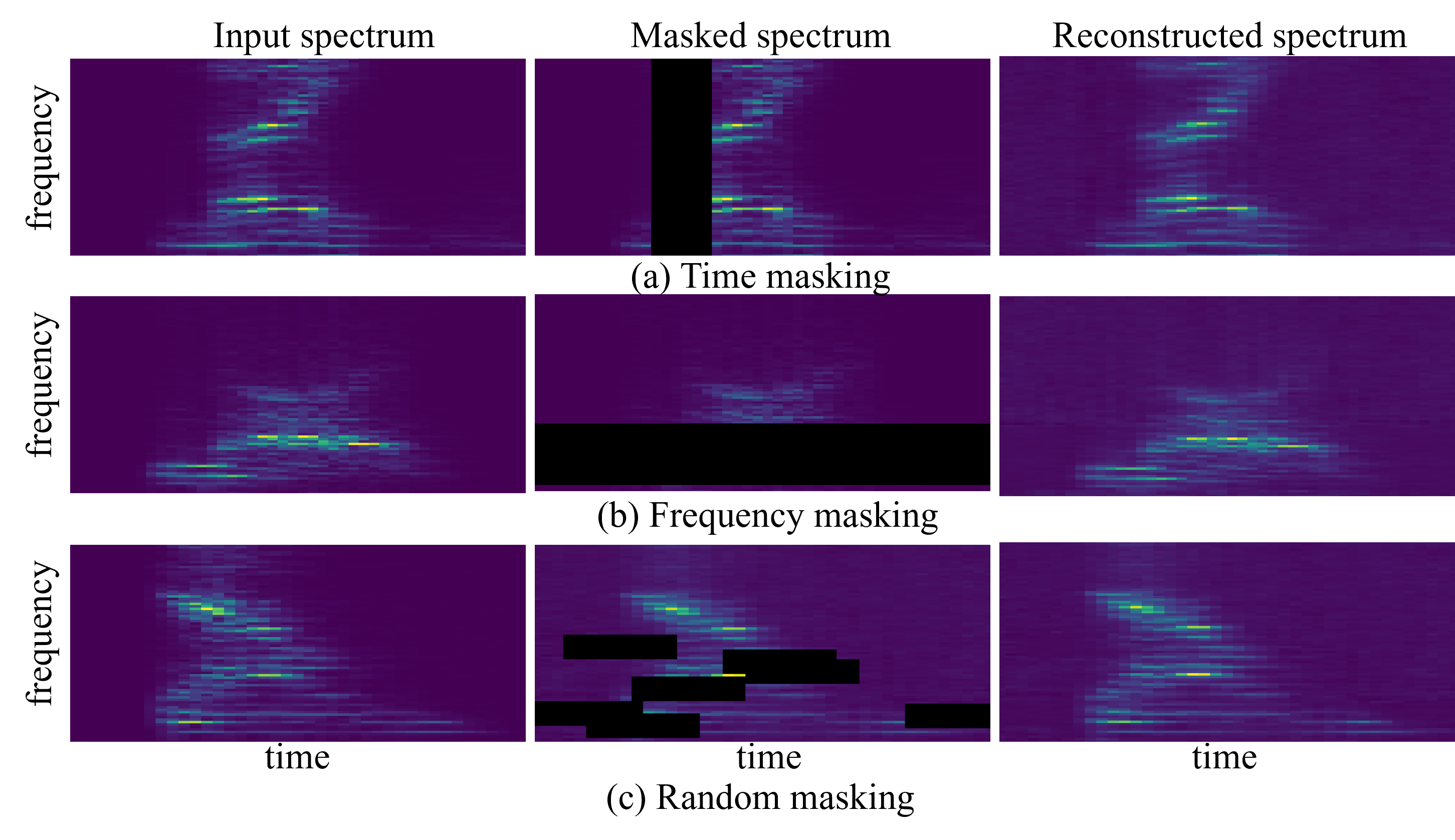}
    \caption{Masking strategy used in the proposed SSCL module and the corresponding reconstructed samples.}
    \label{reconstruction_samples}
\end{figure}

\begin{figure}[t]
    \centering
    \includegraphics[width=0.75\linewidth]{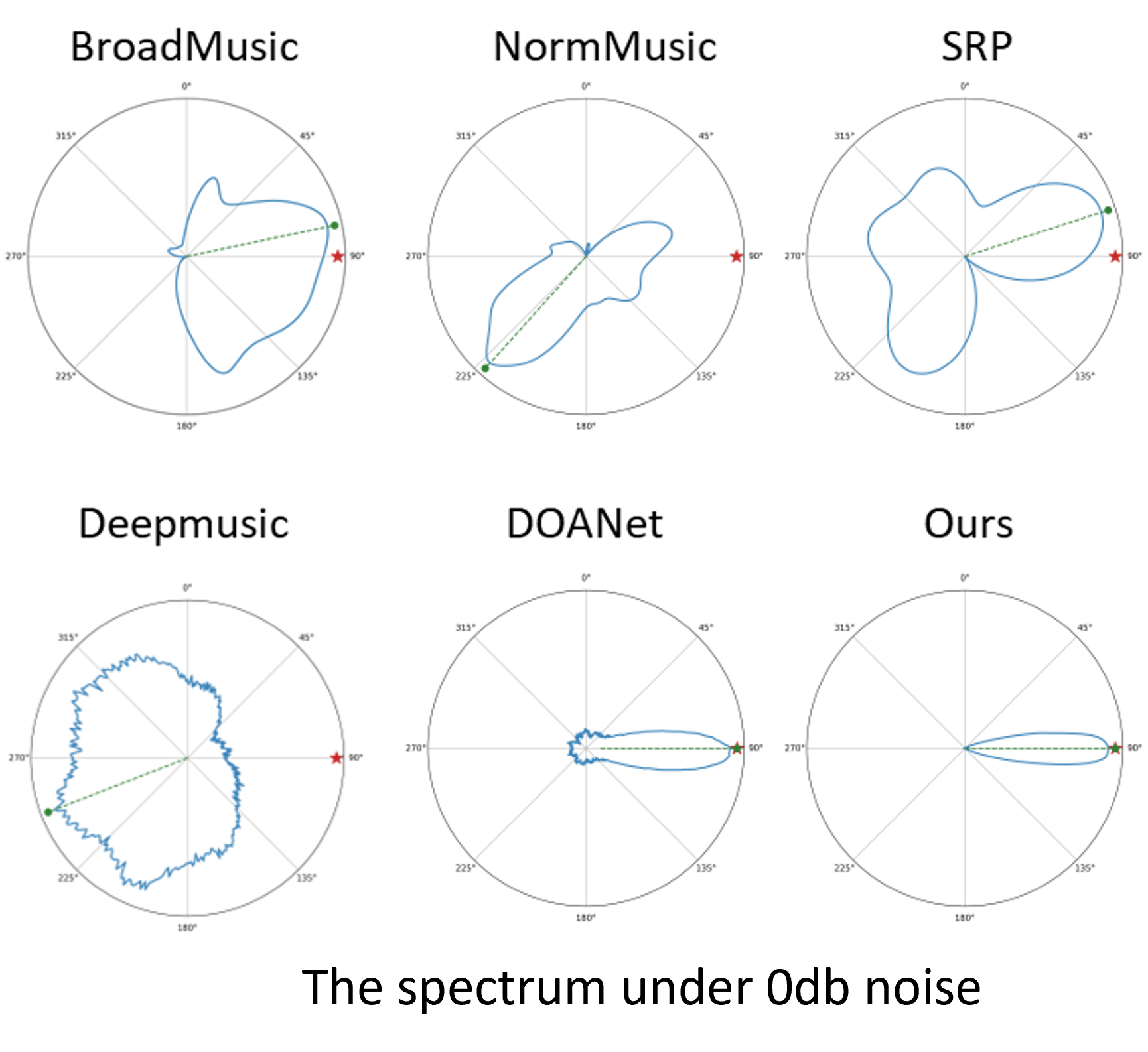}
    \caption{The predicted spectrum for different methods under 0db noise.}
    \label{spectum_noise}
\end{figure}

\subsection{Training Details}

The proposed framework is trained in two stages: 
(1) the {Self-supervised Spatial Correlation Learning} (SSCL) stage and 
(2) the supervised NeuralMUSIC DOA estimation stage.

In the SSCL stage, the model is trained for 400 epochs using the Adam optimizer with a CosineAnnealingLR learning rate scheduler, and a batch size of 256. 
This self-supervised pretraining encourages the network to learn geometry-consistent spatial representations from multi-channel audio, which improves the robustness and generalization capability for subsequent sound source localization tasks.

In the supervised NeuralMUSIC DOA estimation stage, the network is optimized using the Adam optimizer with an initial learning rate of $1\times10^{-3}$ for 100 epochs. 
The balancing parameter $\lambda$ is set to 0.5 and the batch size is 32. 
All deep learning models are implemented in PyTorch and trained on a NVIDIA RTX 4090 GPU.
\begin{figure*}[t]
    \centering
    \includegraphics[width=0.8\linewidth]{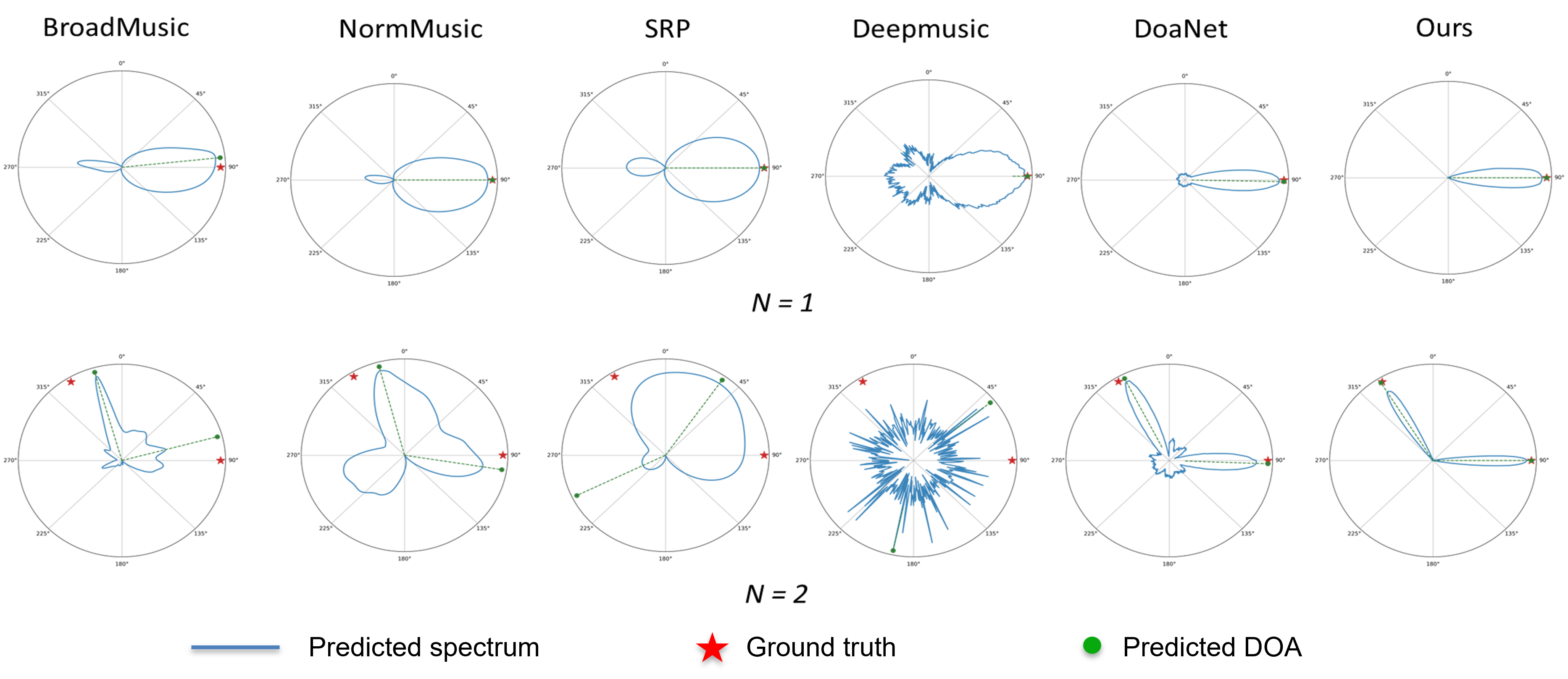}
    \caption{Predicted DOA spectra of different methods under varying numbers of sound sources.}
    \label{spectum_compare}
\end{figure*}

\begin{figure*}[t]
    \centering
    \includegraphics[width=0.9\linewidth]{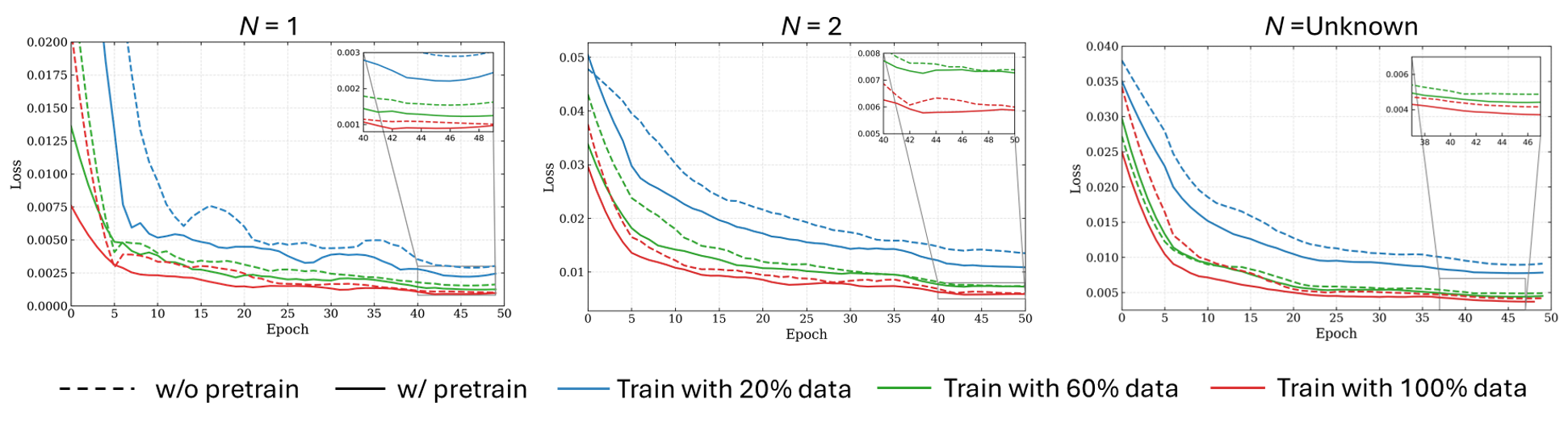}
    \caption{Training loss curve under different training data ratio with (w/) and without (w/o) SSCL}
    \label{train_loss}
\end{figure*}


\subsection{DOA Estimation Results Visualization}

To provide a qualitative comparison of different approaches, we visualize the DOA spectra generated by representative methods. Fig.~\ref{spectum_compare} illustrates the localization results under different sound source number settings. It can be observed that the proposed method produces the most accurate and clean DOA spectra, with sharp peaks located around the ground-truth directions across different scenarios.

In contrast, although traditional subspace-based methods can achieve accurate localization in the single-source case, their performance degrades significantly when multiple sources are present. DOANet \cite{qayyum2020DOANet} also demonstrates competitive performance under both single- and multi-source settings. However, compared with the proposed method, its predicted DOA spectra are less precise, exhibit stronger noise artifacts, and contain more spurious side lobes.

Fig.~\ref{spectum_noise} further presents the localization performance under low signal-to-noise ratio conditions (SNR = 0 dB), where additive white Gaussian noise (AWGN) is introduced into the simulated environment. A similar trend can be observed: the performance of classical methods deteriorates significantly and produces unreliable spectra. In contrast, the proposed method remains robust, generating the cleanest spectral responses and providing the most accurate DOA estimates.

\begin{table*}[ht]
\centering
\caption{Mean Angular Absolute Error (MAAE in $^\circ$) of different methods on the \textbf{SLoClas} and \textbf{AFPILD} datasets under varying training ratios. Lower is better.}
\label{tab:real_dataset}
\renewcommand{\arraystretch}{1.4}
\setlength{\tabcolsep}{8pt}

\begin{tabular}{lcccccccccc}
\toprule
\multirow{2}{*}{\textbf{Method}} 
& \multicolumn{5}{c}{\textbf{SLoClas}} 
& \multicolumn{5}{c}{\textbf{AFPILD}} \\

\cmidrule(lr){2-6} \cmidrule(lr){7-11}

& \textbf{100\%} & \textbf{80\%} & \textbf{60\%} & \textbf{40\%} & \textbf{20\%}
& \textbf{100\%} & \textbf{80\%} & \textbf{60\%} & \textbf{40\%} & \textbf{20\%} \\

\midrule
MUSIC       & 8.93 & 8.93 & 8.93 & 8.93 & 8.93 
           & 16.02 & 16.02 & 16.02 & 16.02 & 16.02 \\

NormMUSIC & 5.96 & 5.96 & 5.96 & 5.96 & 5.96 
           & 15.90 & 15.90 & 15.90 & 15.90 & 15.90 \\

Beamforming & 7.50 & 7.50 & 7.50 & 7.50 & 7.50 
           & 21.49 & 21.49 & 21.49 & 21.49 & 21.49 \\

TOPS      & 16.29 & 16.29 & 16.29 & 16.29 & 16.29 
           & 64.40 & 64.40 & 64.40 & 64.40 & 64.40 \\

CSSM      & 22.88 & 22.88 & 22.88 & 22.88 & 22.88
           & 30.16 & 30.16 & 30.16 & 30.16 & 30.16 \\

WAVES      & 25.18 & 25.18 & 25.18 & 25.18 & 25.18
           & 53.17 & 53.17 & 53.17 & 53.17 & 53.17 \\

FRIDA      & 26.11 & 26.11 & 26.11 & 26.11 & 26.11
           & 56.71 & 56.71 & 56.71 & 56.71 & 56.71 \\

CRNN & 3.37 & 3.87 & 4.72 & 4.78 & 6.71
           & \underline{10.55} & \underline{10.82} & \textbf{10.85} & \underline{11.97} & 22.69 \\

Transformer & \underline{1.77} & \underline{3.31} & \underline{3.56} & \underline{4.51} & \underline{5.32}
            & 11.16 & 11.27 & 12.23 & 12.84 & \underline{14.72} \\

DOANet     & 3.60 & 3.93 & 3.96 & 5.06 & 6.41 
           & 12.78 & 13.09 & 13.73 & 14.32 & 20.00 \\

DeepDAE  & 3.87 & 4.19 & 5.08 & 6.62 & 8.66 
           & 17.29 & 19.48 & 21.98 & 23.00 & 31.63 \\

DeepMusic  & 5.80 & 6.31 & 8.76 & 12.07 & 14.01 
           & 14.98 & 15.04 & 15.87 & 16.73 & 22.86 \\

DA-MUSIC  & 4.87 & 5.79 & 6.60 & 8.28 & 10.29 
           & 15.72 & 17.40 & 17.65 & 20.58 & 29.81 \\

\midrule
\textbf{Ours}
           & \textbf{2.91} & \textbf{3.13} & \textbf{3.33} & \textbf{3.96} & \textbf{5.15}
           & \textbf{10.42} & \textbf{10.48} & \underline{11.22} & \textbf{11.95} & \textbf{13.41} \\

\bottomrule
\label{real_dataset}
\end{tabular}

\end{table*}

\subsection{Effectiveness of SSCL}

In order to better exploit the unlabeled data collected during robot operation, we introduce the Self-supervised Spatial Correlation Learning (SSCL) module. 
The masking strategy used in SSCL and examples of reconstructed samples are illustrated in Fig.~\ref{reconstruction_samples}.

To evaluate whether SSCL benefits the downstream DOA estimation task, we compare the training curves under different data ratios and source-number settings, as shown in Fig.~\ref{train_loss}. 
It can be observed that models initialized with SSCL consistently start from a lower initial loss, indicating a better initialization. 
Moreover, SSCL leads to faster convergence and achieves lower final training losses compared with models trained from scratch. 
This trend remains consistent across different training data ratios and source-number conditions.

These results demonstrate that SSCL provides an effective initialization and enables the network to learn more stable spatial representations, thereby improving both convergence efficiency and overall localization performance in downstream sound source localization tasks.

\subsection{Data Efficiency Testing}
By incorporating physical prior knowledge and effectively leveraging unlabeled data through spatial correlation learning, the proposed method achieves superior data efficiency compared with purely deep-learning-based approaches. 
We further evaluate the proposed method on two real-world datasets. The quantitative results are summarized in Table~\ref{real_dataset}, which further demonstrate the robustness and generalization capability of our approach in practical acoustic environments.

On the \textbf{SLoClas} dataset, the proposed method achieves the lowest MAAE across all training ratios. Even when the training data is reduced to only $20\%$, our method maintains a relatively low error of $5.15^\circ$, while most deep-learning-based methods exhibit noticeable performance degradation.
On the more challenging \textbf{AFPILD} dataset, which involves real-world pedestrian localization scenarios, our method also consistently outperforms existing approaches. In particular, when only $20\%$ of the training data is used, our method achieves an MAAE of $13.41^\circ$, showing strong robustness under limited training data conditions.

Overall, these results demonstrate that the proposed framework achieves superior localization accuracy while maintaining strong robustness and data efficiency across different acoustic environments.

\begin{table}[t]
\centering
\renewcommand{\arraystretch}{1.3}
\setlength{\tabcolsep}{12pt}
\caption{MAAE ($^{\circ}$) of different methods trained using sparse angular sampling (5° and 10° increments) and evaluated on the full DOA range.}
\label{tab:generalization_step}

\begin{tabular}{lcc}
\toprule
\textbf{Methods} & \textbf{Sparse 5° Training} & \textbf{Sparse 10° Training} \\
\midrule

MUSIC   & 3.96 & 3.96 \\
NormMUSIC    & \underline{1.97} & \textbf{1.97} \\
Beamforming  & 2.47 & 2.47 \\
TOPS         & 5.14 & 5.14 \\
WAVES        & 17.04 & 17.04 \\
FRIDA        & 13.01 & 13.01 \\
CRNN    & 7.99 & 9.07 \\
Transformer  & 3.13 & 7.55 \\
DOANet       & 3.22 & 5.77 \\
DeepDAE      & 7.57 & 5.99 \\
DeepMusic    & 3.38 & 14.68 \\
DA-MUSIC     & 6.91 & 12.14 \\

\midrule
\textbf{Ours} & \textbf{1.89} & \underline{2.35} \\

\bottomrule
\end{tabular}

\end{table}

\subsection{Generalization Test on Sparse Training}

To further evaluate the generalization capability of the proposed method, we conduct a more challenging experiment in which the models are trained only on sparsely sampled DOA angles while being evaluated on the full DOA range from $0^\circ$ to $360^\circ$. Specifically, during training, the DOA space is sampled at coarse angular intervals (e.g., $10^\circ$, $20^\circ$, ..., $350^\circ$), whereas the testing phase covers the entire angular space with a $1^\circ$ resolution.

The quantitative results are summarized in Table~\ref{tab:generalization_step}. It can be observed that the proposed method maintains accurate localization performance even when trained with sparse angular supervision. In particular, our method achieves the lowest MAAE under both sparse $5^\circ$ and $10^\circ$ training settings. These results demonstrate that the proposed approach possesses strong interpolation capability and superior generalization compared with existing methods.

\subsection{Future Work}

The proposed NeuralMUSIC framework achieves strong performance in robot sound source localization by integrating data-driven spatial representation learning with classical subspace-based signal processing. By combining physical priors with self-supervised spatial correlation learning, the method demonstrates improved robustness and data efficiency across both simulated and real-world datasets.

Despite these advantages, several limitations remain. First, although the proposed method shows promising generalization ability, it may still underperform traditional signal-processing methods when the training data are extremely limited, when the training distribution is highly imbalanced, or when the microphone array configuration differs significantly from the training setup. Future work will investigate incorporating stronger physical constraints and developing geometry-invariant learning mechanisms to improve cross-array and cross-environment generalization.

Second, further improving data efficiency remains an important direction. Although SSCL helps leverage unlabeled data, the model may still require a certain amount of labeled samples to achieve optimal performance. Future work will explore few-shot learning techniques to enable reliable localization even with only a few labeled samples.

\subsection{Supplementary: Further Discussion on Differences from Existing Hybrid Methods}

Several recent works \cite{shmuel2024subspacenet,merkofer2023music,nguyen2024unet,ji2024transmusic} have explored hybrid frameworks that combine neural networks with classical subspace-based pipelines. 
While these methods demonstrate promising performance, they still exhibit several limitations that restrict their applicability in robotic audition scenarios.

\textbf{(1) Array configuration dependency.}
Some existing approaches are restricted to specific array geometries. 
For instance, works such as \cite{shmuel2024subspacenet,nguyen2024unet} employ Root-MUSIC \cite{FRIEDLANDER199315} within their pipeline, which requires a uniform linear array (ULA). 
Such a constraint limits their applicability to other microphone configurations commonly used in robotic platforms. 
In contrast, our framework is built upon the classical MUSIC formulation, which can be applied to more general array geometries, making it more suitable for diverse robotic sensing setups.

\textbf{(2) Lack of broadband frequency modeling.}
Most existing hybrid methods \cite{shmuel2024subspacenet,merkofer2023music,nguyen2024unet,ji2024transmusic} are designed primarily for narrowband signals, which are common in radar sensing. 
However, acoustic signals in robotic applications are inherently broadband. 
Simply applying narrowband formulations to broadband signals may lead to suboptimal performance because different frequency bands carry varying levels of spatial information. 
If unreliable frequency components are not properly handled, the resulting pseudo-spectrum may be distorted by noise. 
To address this issue, we introduce a Frequency Attention Fusion (FAF) module that adaptively reweights frequency bins and emphasizes informative spectral components when constructing the broadband pseudo-spectrum.

\textbf{(3) Limited validation in real-world environments.}
Many existing hybrid approaches evaluate their performance primarily in simulated environments. 
While simulation provides controlled conditions for algorithm development, real-world acoustic environments are often significantly more complex due to reverberation, sensor noise, and environmental interference. 
Without extensive validation on real-world recordings, the practical effectiveness of these methods remains unclear. 
In contrast, our method is evaluated on three different robotic audio datasets, providing a more comprehensive assessment of its robustness and practical applicability.

\textbf{(4) Loss of interpretability and physical structure.}
Although several hybrid methods combine neural networks with classical algorithms such as MUSIC or Root-MUSIC, these approaches \cite{merkofer2023music,ji2024transmusic} introduce additional neural networks after the classical pipeline to directly regress the DOA. 
Such additional regression modules effectively transform the overall system back into a black-box model and weaken the physical interpretability provided by the subspace formulation. 
This design may also reduce the generalization ability inherited from classical signal processing models. 
In contrast, our framework preserves the classical pseudo-spectrum generation process and introduces only a lightweight frequency attention module to fuse information across frequency bins. 
This design maintains the physical structure of the MUSIC algorithm while improving robustness.

\textbf{(5) Differences in pipeline design.}
Taking DA-MUSIC \cite{merkofer2023music} as an example, its source number estimation module is placed after the eigenvalue decomposition (EVD) stage. 
However, at this stage some information from the raw input may already be lost, as the estimation relies on the intermediate covariance representation. 
This may reduce the accuracy of source number estimation. 
In contrast, our method predicts the source number directly from the input spectral representation, enabling the model to utilize richer spatial information for more reliable estimation.

Furthermore, DA-MUSIC directly regresses the DOA angle. 
This design introduces two potential challenges. 
First, defining an appropriate regression loss becomes more complicated when the number of sources is unknown. 
Second, training stability may be affected because DOA estimation typically involves peak extraction from the pseudo-spectrum, which introduces non-differentiable operations and weak gradients. 
DA-MUSIC addresses this issue by introducing an additional neural regression module. 
While effective, this approach further reduces the interpretability of the overall system. 
In contrast, our method directly outputs the MUSIC pseudo-spectrum and estimates DOA through peak selection. 
This design preserves the physical meaning of the classical algorithm while maintaining a simple and effective learning pipeline. Moreover, representing localization results as a spatial probability distribution rather than a single precise angle is more consistent with human intuition, as it provides an interpretable confidence over possible directions.

\end{document}